%% file: main.tex
\documentclass[journal]{IEEEtran}

\input{settings_preprint}

\usepackage{acro}
\usepackage{enumitem}
\input{sections/sec_acronym}

\ifCLASSINFOpdf
\else
\fi

\begin{document}
%
\title{Using PPP Information to Implement a Global Real-Time Virtual Network DGNSS Approach}
%
%
%

\author{Wang~Hu,
	    Ashim~Neupane, 
        and Jay~A.~Farrell,~\IEEEmembership{Fellow,~IEEE}
\thanks{Copyright (c) 2022 IEEE. Personal use of this material is permitted. However, permission to use this material for any other purposes must be obtained from the IEEE by sending a request to pubs-permissions@ieee.org. Manuscript received 11/17/2021; revised 4/11/2021, accepted 6/24/2022.}
\thanks{W. Hu, A.~Neupane, and J. A. Farrell are with the Department of Electrical and Computer Engineering,
	University of California, Riverside, CA 92521 USA (e-mail: \{whu027, aneup001, farrell\}@ucr.edu).}%
\thanks{This work was supported by Caltrans under agreement number 65A0767.}
}

%
%

\markboth{Preprint: Please cite final version in IEEE Transactions on Vehicular Technology}%
{Hu \MakeLowercase{\textit{et al.}}: A Global Real-Time Virtual Network DGNSS Approach}
%
%



\maketitle

\begin{abstract}
\ac{DGNSS} has been demonstrated to provide reliable, high-quality range correction information enabling  real-time navigation with centimeter to  sub-meter accuracy, which is required for applications such as  connected and autonomous vehicles. 
However, \ac{DGNSS} requires a local reference station near each user.
For a continental or global scale implementation, this information dissemination approach would require a dense network of reference stations whose construction and maintenance would be prohibitively expensive. 
\ac{PPP} affords more flexibility as a public service for \ac{GNSS} receivers, but its \ac{SSR} format is not supported by most receivers in the field or on the market. 

This article proposes a novel \ac{VNDGNSS} approach and an optimization algorithm that is key to its implementation.
The approach capitalizes on the existing \ac{PPP} infrastructure without the need for new physical reference stations.
Specifically, no reference station is needed in the local vicinity of any user. 
By connecting to public \ac{GNSS} \ac{SSR} data services, a \ac{VNDGNSS} server maintains current information about satellite code bias, satellite orbit and clock, and atmospheric models.
Construction of the \ac{RTCM} \ac{OSR} messages from this \ac{SSR} information requires both the signal time-of-transmission and the satellite position at that time which are consistent with the time-of-reception for each client. 
This article presents an algorithm to determine these quantities.
Then the \ac{VNDGNSS} computes and transmits \ac{RTCM} \ac{OSR} messages to user receivers. 
This approach achieves global dissemination coverage for real-time navigation without the need for additional local base stations near each user.

The results of real-time stationary and moving platform evaluations are included, using u-blox M8P and ZED-F9P receivers.
The performance surpasses the Society of Automotive Engineering (SAE) specification (68\% of horizontal error $\leqslant$ 1.5 m and vertical error $\leqslant$ 3 m) and shows significantly better horizontal performance than GNSS \ac{OS}. 
The moving tests also show better horizontal performance than the ZED-F9P receiver with \ac{SBAS} enabled and achieve the lane-level accuracy (95\% of horizontal errors less than 1 meter).
\end{abstract}


%
\IEEEpeerreviewmaketitle

\section{Introduction} 
\input{sections/sec_intro}

%
%
%
%


\section{GNSS Pseudorange Observations} \label{sec:gnssmodel}  
\input{sections/sec_gnssmodel}

\section{Satellite Position, Range, and Time-of-Transmission} \label{sec:satpos}

\input{sections/sec_satposest}

\section{Common-mode Error Calculations} \label{sec:comcorr}
This  section focuses on the use of \ac{SSR} data to compute the common mode error terms in the second line of Eqn. (\ref{eqn:RTCM}).
These calculations will depend on the given base position $P_b$ which is considered as known to the server 
and 
the satellite position $\hat{P}^s(t^s)$ which is computed as described in Section \ref{sec:satpos}.

\subsection{Satellite Orbit and Clock Corrections}
\label{sect:orbit_clock}
\input{sections/subsec_satobtclkcorr}

\subsection{Satellite Hardware Bias}

\input{sections/subsec_codebias}

\subsection{Global Troposphere Model} \label{sec:trop}
\input{sections/subsec_trop}

\subsection{Ionosphere Model} \label{sec:iono}

\input{sections/subsec_iono}

\subsection{Discussion} 
\label{sect:disc1}
The present server approach is designed for pseudorange positioning, using manufacturer software. 
It was not designed for carrier phase (i.e., \ac{RTK}) positioning for a few reasons.
First, the project supporting this research was focused on single-frequency approaches, while
practical PPP-RTK requires a multi-frequency solution to allow widelane implementations (Sec. 21.4.5 in \cite{teunissen2017springer}).
Second, the use of a virtual base instead of a physical base means that various of the common-mode errors (e.g., satellite hardware biases) are not accounted for (in the corrections) to the extent required for the double-difference operations to result in `integer' ambiguities \cite{PsiakiMohiuddin2012}.

\input{sections/sec_newdesign}

\section{Experimental  Performance Analysis} \label{sec:test}
\input{sections/sec_expresults}

\input{sections/sec_conclusion}

\section*{Acknowledgments}
The ideas reported herein originated during a project supported by SiriusXM as reported in \cite{rahman2019ecef}. 
This research support is greatly appreciated.
The ideas reported herein, and any errors or  omissions, are the responsibility of the authors and do not reflect the opinions of the sponsors.

\ifCLASSOPTIONcaptionsoff
  \newpage
\fi



%

\bibliographystyle{biblio/IEEEtran}
\bibliography{biblio/IEEEabrv,biblio/References}

\input{biography/Biography}




\clearpage
\appendix
The following material is included for the benefit of the reviewers only. 
It would not appear in the published version.

\subsection{Derivation of Partial Derivative}
\label{app:partial}

The objective of this section is to derive Eqn. (\ref{eqn:partial1}).

For the error function $F({t}_p^s)$ defined in Eqn. (\ref{eqn:F_defn}),
the partial derivative is evaluated as follows:
\begin{align}\label{eqn:partial}
	\frac{\partial F}{\partial {t}_p^s}&=
	\frac{\partial }{\partial {t}_p^s}\left(R\Big(\BoldP_b, \hat{\BoldP}^s({t}_b - {t}_g^s)\Big) 
	- c \, {t}_g^s \right)
	\nonumber \\
	&= \left(
	\frac{\partial }{\partial {t}_g^s} R\Big(\BoldP_b, \hat{\BoldP}^s({t}_b - {t}_g^s)\Big) 
	- c \right) \frac{\partial {t}_g^s}{\partial {t}_p^s}
	\nonumber \\
	&=-\,\left(\left.
	\frac{\partial }{\partial {\BoldP}^s} 
	R\Big(\BoldP_b, {\BoldP}^s\Big)\right|_{{\BoldP}^s=\hat{\BoldP}^s({t}_b - {t}_g^s)}
	\left.
	\frac{\partial {\BoldP}^s}{\partial t}  
	\right|_{t={t}_b - {t}_g^s}
	+c  \right) \frac{\partial {t}_g^s}{\partial {t}_p^s}
	\nonumber  
\end{align}  
where 
$\1_b^s = \left.
\frac{\partial }{\partial {\BoldP}^s} 
R\Big(\BoldP_b, {\BoldP}^s\Big)\right|_{{\BoldP}^s=\hat{\BoldP}^s({t}_b - {t}_p^s)}$ is a unit vector 
and 
$$\hat{\BoldV}^s({t}_b - {t}_p^s) = \left.\frac{\partial {\BoldP}^s}{\partial t}  
\right|_{t={t}_b - {t}_p^s}.$$
Given that ${t}_g^s = {t}_p^s + \hat{dt}^s(\hat{t}^s)$ and $\hat{t}^s = t_b - {t}_p^s$,
\begin{align} 
	\frac{\partial {t}_g^s}{\partial {t}_p^s}
	&= 1 + \frac{\partial \hat{dt}^s(\hat{t}^s)}{\partial {t}_p^s}   \\
	&= 1 + \left(\frac{\partial }{\partial \hat{t}^s}\hat{dt}^s(\hat{t}^s) \right)~\left(
	\frac{\partial \hat{t}^s}{\partial {t}_p^s}  \right) \\
	&= \left(1 - \frac{\partial }{\partial \hat{t}^s} \hat{dt}^s(\hat{t}^s)~
	\right) \\
	&=\left(1 - ( a_1 + 2\,a_2\,(\hat{t}^s-t_{oc}))  \right),
\end{align}
where $a_1$, $a_2$, and $t_{oc}$ are the clock model parameters from the navigation message. 
Therefore, 
\begin{align}
	\frac{\partial F}{\partial {t}_p^s}
	&= -\left( \1_b^s \, \cdot \, \hat{\BoldV}^s({t}_b + {t}_g^s) 	+ c  \right)
	\left(1 - a_1 - 2\,a_2\,(\hat{t}^s-t_{oc})  \right).
	\nonumber
\end{align} 

\end{document}

%% file: settings_preprint.tex
\usepackage{graphics}
\usepackage{graphicx} 
\usepackage{multirow}
\usepackage{longtable}
\usepackage{epsfig}
\usepackage{epstopdf}

\usepackage{amsmath, mathrsfs, dsfont}
\usepackage{amssymb}
\usepackage{amsfonts}
\usepackage{bm}

\usepackage{cite}
\usepackage{verbatim}

\usepackage{amscd}
\usepackage{color}
\definecolor{lgray}{gray}{0.6}
\usepackage{rotating}

\usepackage[font=footnotesize]{subcaption}

\usepackage{mathptmx}
\usepackage[11pt]{moresize}
\usepackage{flushend}
\usepackage[stable]{footmisc}

\usepackage{algorithmicx}
\usepackage{algorithm}
\usepackage{algpascal}
\usepackage{float}
\usepackage{algc}
\usepackage{algcompatible}
\usepackage{algpseudocode}

\usepackage[vcentermath]{youngtab} 
\usepackage{url}

\newcommand{\Bolda}				{ \mathbf{a} }

\newcommand{\BoldI}				{ \mathbf{I} }

\newcommand{\BoldO}				{ \mathbf{O} }
\newcommand{\BoldP}				{ \mathbf{P} }

\newcommand{\BoldR}				{ \mathbf{R} }

\newcommand{\BoldS}				{ \mathbf{S} }

\newcommand{\BoldV}				{ \mathbf{V} }

\newcommand{\Boldw}				{ \mathbf{w} }

\newcommand{\1}					{ \boldsymbol{1} }

\newcommand{\Boldtheta}			{ \boldsymbol{\theta} }

\newcommand\T{\rule{0pt}{2.6ex}}       
\newcommand\B{\rule[-1.2ex]{0pt}{0pt}} 

\newcounter{inlineenum}
\renewcommand{\theinlineenum}{\alph{inlineenum}}

	\newcommand{\brmrk}[1]{\begin{remark} \label{#1} }
	\newcommand{\ermrk}{ \hfill $\bigtriangleup$    \end{remark} \vspace{1mm} }

\newtheorem{exercise}{Exercise}[section]
\newcommand{\boex}[1]{\begin{example} \label{#1} --- \rm}
	\newcommand{\eoex}{ \hfill $\bigtriangleup$    \end{example} \vspace{1mm} }
\newtheorem{example}{Example}[section]
\newcommand{\bohw}[1]{\begin{exercise} \label{#1} -- \rm}
	\newcommand{\eohw}{ \hfill    \end{exercise} \vspace{1mm} }
\newtheorem{assumption}{Assumption}[section]
	\newcommand{\boass}[1]{\begin{assumption} \label{#1} -- \rm}
	\newcommand{\eoass}{ \hfill    \end{assumption} \vspace{1mm} }

\renewcommand{\baselinestretch}{1.0}

\newcommand{\black}{\color{black}}
\newcommand{\blue}{\color{blue}}

\definecolor{brinkpink}{rgb}{1.00, 0.33, 0.64}

%% file: sections/sec_acronym.tex
\DeclareAcronym{APC}{
	short=APC,
	long=Antenna Phase Centre,
}

\DeclareAcronym{COM}{
short=COM,
long=Center of Mass,
}

\DeclareAcronym{CAV}{
short=CAV,
long=Connected and Autonomous Vehicle,
}

\DeclareAcronym{DCB}{
short=DCB,
long=Differential Code Bias,
}
\DeclareAcronym{CDMA}{
short=CDMA,
long=Code Division Multiple Access,
}

\DeclareAcronym{FDMA}{
	short=FDMA,
	long=Frequency Dvision Multiple Access,
}
\DeclareAcronym{DGPS}{
	short=DGPS,
	long=Differential Global Positioning System,
}
\DeclareAcronym{DGNSS}{
short=DGNSS,
long=Differential GNSS,
}
\DeclareAcronym{ECEF}{
short=ECEF,
long=Earth-Centered Earth-Fixed,
}
\DeclareAcronym{GPS}{
	short=GPS,
	long=Global Positioning System,
}
\DeclareAcronym{GNSS}{
short=GNSS,
long=Global Navigation Satellite Systems,
}
\DeclareAcronym{GPSSPS}{
short=GPS SPS,
long=GPS standard positioning service,
}

\DeclareAcronym{OS}{
short=OS,
long=Open Service,
}
\DeclareAcronym{OSB}{
short=OSB,
long=Observable-specific Code Biases,
}
\DeclareAcronym{OSR}{
short=OSR,
long=Observation Space Representation,
}
\DeclareAcronym{PPP}{
	short=PPP,
	long=Precise Point Positioning,
} 
\DeclareAcronym{PPP-AR}{
short=PPP-AR,
long=Precise Point Positioning Ambiguity Resolution,
} 
\DeclareAcronym{RTCM}{
short=RTCM,
long=Radio Technical Commission for Maritime Services,
}
\DeclareAcronym{RTK}{
short=RTK,
long=Real-time Kinematic Positioning,
}
\DeclareAcronym{SBAS}{
short=SBAS,
long=Satellite Based Augmentation Systems,
}

\DeclareAcronym{SNR}{
short=SNR,
long=Signal-to-Noise Ratio,
}
\DeclareAcronym{SSR}{
short=SSR,
long=State Space Representation,
}
\DeclareAcronym{SPS}{
	short=SPS,
	long=Standard Positioning Service,
}
\DeclareAcronym{STEC}{
	short=STEC,
	long=Slant Total Electron Content,
}
\DeclareAcronym{VTEC}{
	short=VTEC,
	long=Vertical Total Electron Content,
}

\DeclareAcronym{SH}{
	short=SH,
	long=Spherical Harmonic,
}
\DeclareAcronym{TGD}{
short=TGD,
long=Timing Goup Delay,
}
\DeclareAcronym{ZTD}{
short=ZTD,
long=Zenith Troposphere Delay,
}
\DeclareAcronym{TEC}{
short=TEC,
long=Total Electron Content,
}
\DeclareAcronym{IPP}{
	short=IPP,
	long=Ionosphere Pierce Point,
}
\DeclareAcronym{NOAA}{
	short=NOAA,
	long=National Oceanic and Atmospheric Administration,
}
\DeclareAcronym{UCR}{
	short=UCR,
	long=University of California-Riverside,
}
\DeclareAcronym{USTEC}{
short=US-TEC,
long=US Total Electron Content,
}
\DeclareAcronym{VNDGNSS}{
	short=VN-DGNSS,
	long=Virtual Network DGNSS,
}
\DeclareAcronym{IOD}{
	short=IOD,
	long=Issue Of Data,
}
\DeclareAcronym{CAS}{
	short=CAS,
	long=Chinese Academy of Sciences,
}
\DeclareAcronym{CNES}{
	short=CNES,
	long=Centre national d'études spatiales,
}
\DeclareAcronym{RMS}{
	short=RMS,
	long=Root Mean Square,
}
\DeclareAcronym{SF}{
	short=SF,
	long=Single Frequency,
}
\DeclareAcronym{DF}{
	short=DF,
	long=Dual Frequency,
}
\DeclareAcronym{ICD}{
	short=ICD,
	long=Interface Control Document,
}
\DeclareAcronym{IGS}{
	short=IGS,
	long=International GNSS Service,
}
\DeclareAcronym{NED}{
	short=NED,
	long={North, East and Down},
}
\DeclareAcronym{WAAS}{
	short=WAAS,
	long=Wide Area Augmentation System,
}
\DeclareAcronym{OPUS}{
	short=OPUS,
	long=Online Positioning User Service,
}

\DeclareAcronym{VRS}{
	short=VRS,
	long=Virtual Reference Station,
}

\DeclareAcronym{VBS}{
	short=VBS,
	long=Virtual Base Station,
}

\DeclareAcronym{SPP}{
	short=SPP,
	long=Single-frequency Point Positioning,
}

\DeclareAcronym{BNC}{
	short=BNC,
	long=BKG NTRIP Client,
}

\DeclareAcronym{ITS}{
	short=ITS,
	long=Intelligent Transportation System
}

%% file: sections/sec_intro.tex
\IEEEPARstart{I}{mportant} 
real-time  transportation  applications require positioning services with high accuracy and reliability (e.g., lane determination, lane-level queue determination, lane-specific direction guidance or traffic signal management, eco driving, speed harmonization)\cite{sae2016,farrell2012its,NigelBarth2021,barth2012eco}. 
\ac{GNSS} is a popular potential solution approach. 

The \ac{GNSS} \ac{OS} without external correction service typically achieves 10 meters accuracy in real-time under the open sky. 
Specifically, GPS, Galileo, and BeiDou, respectively, provide 
95\% of horizontal error less than 8 m, 7.5 m, and 10 m; 
95\% of vertical error less than 13 m, 15 m, and 10 m \cite{team2014global,beidou2018os,galileo2019os}. 
This \ac{OS} accuracy is not sufficient to satisfy the requirements of many
 \ac{ITS} and \ac{CAV} 
applications.
Several references state position accuracy specifications, for instance: 
the Society of Automotive Engineering (SAE) specification requires 68\% of horizontal error to be less than 1.5 m and vertical error to be less than 3 m \cite{sae2016}; and, 
lane-level positioning accuracy for \ac{ITS} applications  require 95\% of horizontal error to be less than 1 m \cite{NigelBarth2021,farrell2012its}.
Mitigation of common-mode \ac{GNSS} measurement errors (e.g., satellite orbit and clock errors, hardware bias, atmospheric delay) is required to improve position accuracy in real-time navigation to achieve these specifications.

Two \ac{GNSS} correction approaches have been demonstrated to effectively improve the \ac{GNSS} positioning accuracy: \ac{OSR} and \ac{SSR}. 
Traditional \ac{DGNSS} approaches use \ac{OSR} corrections, for example, a local base station sending its measurements in an \ac{RTCM} format allows correction of common-mode errors by nearby receivers.
Code-measurement \ac{DGNSS} can achieve 0-3 meters accuracy, while the phase-measurement \ac{DGNSS} (i.e., \ac{RTK}), can achieve centimeter-level accuracy when the carrier phase integer ambiguities are correctly resolved (see Table 21.7 of \cite{teunissen2017springer}). 
In \ac{OSR} approaches, each receiver must change its base station when its motion carries it beyond the coverage area of the current base station.
\ac{PPP} approaches  \cite{rahman2019ecef} rely on correction information communicated in \ac{SSR} format.
Different portions of the \ac{SSR} messages correct each type of common-mode  error.  
The global accuracy of the information in the \ac{SSR} messages is maintained and distributed freely by public governmental agencies, using existing worldwide networks of \ac{GNSS} receivers.
While most commercially available receivers do accept \ac{RTCM} \ac{OSR} messages, they do not currently accept \ac{RTCM} \ac{SSR} messages. 
Even after the next generations of receivers begin to accept \ac{SSR} messages, there will still be numerous previously installed receivers that can only use \ac{RTCM} \ac{OSR} messages.

There are approaches currently available that enable correction of common-mode errors using \ac{PPP} type information in commercial receivers, under limited circumstances. 
Commercial services, such as Trimble RTX,  use \ac{PPP} technology to provide real-time, centimeter-level positioning worldwide. 
The corrections are delivered through cellular or their communication satellite.
This service is only available for specific Trimble receivers at a high service cost \cite{chen2011trimble} and their technical approach is not described in the open literature.
\ac{SBAS} services (e.g., WAAS for GPS in North America, EGNOS for Galileo in Europe, and BDSBAS for BeiDou in China) offer increased accuracy over wide-areas, approaching continental scale, for their specific \ac{GNSS} system. 
However, there are limitations of the existing \ac{SBAS}. 
First, many legacy receivers are not compatible with \ac{SBAS} or \ac{PPP}.
Second, groups of nations have implemented forms of \ac{SBAS} within limited geographic regions. 
At present, there is no globally available SBAS. 
Even within the area that an SBAS is specified to operate, its SBAS information is only available when that system's SBAS satellite is tracked.
Third, each SBAS provides corrections to only one satellite constellation, which limits the performance robustness. 
For example, within North American, WAAS supplies corrections for GPS, but not for Galileo, Beido, or GLONASS. 
Even within North America, terrain, buildings, or foliage may limit the number of GPS satellites that are visible.
Satellites from other constellations would still be available, but without corrections; therefore, accuracy will be degraded relative to the \ac{VNDGNSS} approach described herein that can provide corrections for  Multi-GNSS  satellites.

Existing Network DGNSS approaches also aim to overcome the limitations of local DGNSS.
The \ac{VRS} or \ac{VBS} approaches \cite{vollath2000multi,wanninger1998real} use observations from a network of reference stations.
A master station virtually shifts actual observations from the reference stations to the rover site. 
These approaches offer wide-area coverage, determined by the extent of the reference station network. 
Due to the cost of installing and maintaining a reference station network over a large geographic region, existing large-scale implementations are mainly commercial and fee-based.
Trimble VRS, supporting GPS and GLONASS, claims to provide centimeter-level accuracy for parts of the area of North America, Europe, and Australia \cite{trimble_vrs,chen2011trimble}. 
This service is only supported by specific Trimble receivers.
OmniSTAR VBS is a GPS L1 only, code phase pseudo-range solution on a global scale. 
It generates \ac{RTCM} SC104 \ac{OSR} corrections for the user specified location. 
This service claims that a typical 24-hour sample will show 95\% with less than 1-meter horizontal error \cite{omni_vbs}.
Both Trimble VRS and OmniSTAR VBS are subscription based services.
Qianxun SI FindM is another sub-meter accuracy positioning service, but it is only available within China \cite{qianxun_findm}.
The methodology of these approaches is not reported in the open literature.
These Network \ac{DGNSS} approaches rely directly on receiver's measurements, not \ac{PPP} information. 

This paper proposes an open-source\footnote{The VN-DGNSS repository is public on GitHub: \url{https://github.com/Azurehappen/Virtual-Network-DGNSS-Project}.
	\label{ftnt:GitHub}} that uses publicly available, real-time, PPP data and error models to construct
 \ac{RTCM} \ac{OSR} messages at the user-specified location.
This approach can provide a Multi-GNSS correction service to users globally, without cost to the user.
The \ac{RTCM} \ac{OSR} messages that are transmitted are accepted by most receivers. 
In this approach, there is no need for additional {\em physical} reference stations. 
The \ac{RTCM} \ac{OSR} format communicates corrections in the form of an observation, which must be computationally constructed in any \ac{VNDGNSS} approach using PPP information. 
The computationally constructed observation requires the satellite position and time-of-transmissions for each satellite to be consistent for measurements at a specified time at the user-specified location.  
Section \ref{sec:satpos}  presents an optimization algorithm that solves this problem.
Section \ref{sec:comcorr} describes the use of PPP information to construct the observation.
Section \ref{sec:NewDes} discusses the client-server architecture and PPP information sources.
The currently implemented system
is single frequency (i.e., GPS L1, Galileo E1, and BeiDou B1) and pseudorange only.
Section \ref{sec:test} describes experimental results.
Our initial tests demonstrate the feasibility of improving positioning accuracy to surpass the SAE specification. 
The moving tests also achieve lane-level accuracy requirements.
Its extension to carrier phase and multi-frequency are discussed in Sections \ref{sect:disc1} and \ref{sect:concl}, respectively.

%% file: sections/sec_gnssmodel.tex
	Each \ac{GNSS} constellation $\BoldS$, that uses \ac{CDMA}\footnote{The signals transmitted by GPS, Galileo and BeiDou are based on  \ac{CDMA} \cite{icd2013global,galileo2008galileo,beidou2012beidou}. 
	GLONASS is based on  \ac{FDMA}\black \cite{glonass2008glonass}. 
	For constellations based on CDMA, the frequencies are identical for all satellites.
	The measurement models presented in this section and the construction of this article are based on CDMA. 
	For constellations based on \ac{FDMA}, the frequency is different for each satellite and therefore for each receiver channel; therefore, there are additional frequency dependent \emph{inter-channel biases} (ICBs) terms in both code and carrier phase observation equations \cite{song2014impact,wan2012carrier}.\label{ftnt:FDMA}}, transmits signals using $f$ carrier frequencies. 
	The code (pseudorange) observation from satellite $s$ tracked by receiver $r$ at the GNSS  receiver time $t_r$  is modeled as (see Chap. 21 of \cite{teunissen2017springer}) 
	\begin{equation}\label{eqn:codeM}
		\left.
		\begin{aligned}
			\rho^s_{f,r} (t_r) = &R\Big(\BoldP_r(t_r), \hat{\BoldP}^s(t^s)\Big) 
			+ E_r^s(t^s) 
			- c \,  dt^s 
			+ c \, B_f^s \\
			& + c \, dt_r + c \,  d_r^S  
			+ T^s_r + I^s_{f,r} + M^s_{f,r} + \eta^s_{f,r},
		\end{aligned}
	\right\}
	\end{equation}
	where 
	$\BoldP_r$ denotes the receiver antenna position, 
	$\hat{\BoldP}^s$ denotes the satellite position computed using the satellite navigation message, 
	while $t^s$ denotes the time when  the signal was transmitted. 
	The symbols $dt^s$ and  $\hat{dt}^s$, and other important symbols,  are defined as in Table \ref{tbl:notation}.
	\black
	
\begin{table}[bt]
	\centering
	\caption{Notation definitions.}
	\label{tbl:notation}
	\begin{tabular}{rl}	
		$\omega_{ie}$: 	& {Earth rotational rate (rad/s)},\\
		$c$: 			& {Speed of light (m/s)},\\
		$E^s_r$: 		& {Satellite range ephemeris error (m)},\\ 
		$\delta t^s$:  	& {Satellite clock model error, 	$\delta t^s =(\hat{dt}^s -dt^s)$,  (s)},\\
		$\hat{dt}^s$: 	& {Broadcast satellite clock correction  (s)},\\
		$dt^s$: 		& {Satellite clock bias (s)},\\
		$B_f^s$: 		& {Satellite code hardware bias (s)}, \\
		$dt_r$: 		& {Receiver clock error (s)}, \\
		$d_r^S$:  		& {Receiver code hardware bias for system S (s)},\\ 
		$T^s_r$: 		& {Tropospheric delay (m)},\\
		$I^s_{f,r}$: 	& {Ionospheric delay for $f$ frequency (m)},\\
		$\eta^s_{f,r}$:	& {Random code measurement noise (m)},\\
		$M^s_{f,r}$: 	& {Multipath error (m)}.
	\end{tabular}
\end{table}	
	
	Both $\BoldP_r$ and $\hat{\BoldP}^s$ are defined in the \ac{ECEF} frame.
	Because the \ac{ECEF} frame is rotating, the range quantity in Eqn. (\ref{eqn:codeM}) must be interpreted carefully, as the \ac{ECEF} frame is oriented differently at times $t^s$ and $t_r$.	
	Let $E_r$ and $E_s$ represent the \ac{ECEF} frames at time $t_r$ and $t_s$ respectively.
	The rotation matrix $R^{E_r}_{E_s}$ rotates vectors from their representation in frame $E_s$ to $E_r$ 
	Therefore, $$R(\BoldP_r(t_r), \hat{\BoldP}^s(t^s)) = \|\BoldP_r(t_r)-\BoldR^{E_r}_{E_s}~\hat{\BoldP}^s(t^s)\|.$$
	Alternatively, this range can be computed by the well known Sagnac correction:
	\begin{equation} \label{eqn:Rsag}
	R(\BoldP_r(t_r), \hat{\BoldP}^s(t^s)) = \|\BoldP_r(t_r)-\hat{\BoldP}^s(t^s)\| + \frac{\omega_{ie}}{c}(xb-ay),
	\end{equation}
	where $\BoldP^s = [x,y,z]^T$ and $\BoldP_r = [a,b,c]^T$.
	The accuracy of the Sagnac correction is analyzed in \cite{hu2019derivation} where it is shown to be accurate to $10^{-6}$ m.

	The goal of the state estimator using GNSS is to accurately estimate the state vector, 
	which will include the position $\BoldP_r$, 
	the receiver hardware bias for each system $d_r^S$,  and 
	the receiver clock error $\delta t_r$.
	The  model for $\hat{dt}^s$,  which includes a polynomial and  a relativistic correction, can be found in the \ac{ICD} of each constellation (Sec. 20.3.3.3.3.1 in \cite{icd2013global}, Sec. 5.1.4 in \cite{galileo2008galileo}, and Sec. 7.4 in \cite{beidou2012beidou}).
	The remaining satellite clock model error is 	
	$\delta t^s =(\hat{dt}^s -dt^s)$.
	
	The error terms of the measurement model of Eqn. (\ref{eqn:codeM})  can be classified into two categories:
	\begin{itemize}
		\item Common-mode errors are essentially the same for all receivers in a local vicinity. These include: 
		satellite range ephemeris and clock model errors, 
		satellite code hardware bias, and
		atmospheric (ionospheric and tropospheric) error.   
		\item Non-common-mode errors are different for each receiver. These include: 
		code multipath error and
		random code  measurement noise. 
	\end{itemize}
	These errors significantly affect the positioning accuracy. 
	The typical error magnitudes are summarized in Table. \ref{tab:error}.
	
	\begin{table}[bt]
		\centering
		\caption{GNSS range measurement error magnitudes.}
		\begin{tabular}{c|c|c}
			\hline
			Error Source      & Typical Magnitude   & Reference   \T                                            \\ \hline
			Satellite range ephemeris   & $\sim$ 2 m  & \cite{bryan2018eph}           \T      \\ \hline
			Satellite clock   & $\sim$ 5 ns & \cite{bryan2018eph}      \T       \\ \hline
			Satellite code bias         & $\sim$ 20 ns & \cite{geng2019modified,wang2016deter,mont2014diff}                 \T   \\ \hline
			Troposphere delay & $\sim$ 2.3 m & \cite{bock2001atm} \T  \\ \hline
			Ionosphere delay  & \textless\,30 m & Sec. 25.2 in \cite{teunissen2017springer}  \T\\ \hline
			Code multipath   & \textless\,3 m  &  Sec. 8.4.7 in \cite{farrell2008aided}          \T  \\ \hline
			Measurement noise & \textless\,1 mm  & Sec. 19.7.1 in  \cite{teunissen2017springer}   \T  \\ \hline
		\end{tabular}
		\label{tab:error}
	\end{table}

	There are three standard methods to decrease the effects of GNSS measurement errors. 
	First, for GNSS \ac{OS} approaches (i.e.,  \ac{GPSSPS}, Galileo \ac{OS}, and Beidou \ac{OS}), 
	 the \ac{SPP} solution only uses a prior troposphere model and an ionosphere model with parameters from the navigation message to reduce the atmospheric errors (Chap. 21 of \cite{teunissen2017springer}).
	 GNSS OS approaches are still affected by  satellite biases, residual ephemeris and  clock errors.
	Second, for \ac{OSR} approaches using the  \ac{RTCM} standard, 
	GNSS measurements from a receiver at a known location near to the user receiver are communicated to the user receiver, which forms differential measurements that are essentially free from common-mode errors \cite{rahman2019ecef}.
	Essentially all receivers can accept RTCM OSR messages.
	Third, \ac{SSR} approaches communicate correction information for each portion of the common-mode range errors separately \cite{laurichesse2010real,geng2019modified} from an on-line data source to the receiver.
	At present, few receivers are available that accept \ac{SSR} correction information.
	Even as more receivers become available that accept \ac{SSR} correction information, legacy receivers will require correction information in \ac{OSR} format. 
	This paper presents a method to compute OSR format differential information from on-line \ac{SSR} information, which alleviates the need for local base stations. 
	The \ac{RTCM} format for communicating \ac{OSR} correction information is as a measurement.

	Considering Eqn. (\ref{eqn:codeM}) from the perspective of constructing an \ac{RTCM} \ac{OSR} message using \ac{SSR} data for  a known virtual base antenna position $\BoldP_b$. 
	First, the base station receiver clock error $\Delta t_b^S =  dt_b + d_b^S$ is not important for positioning applications.  
	This  error will only affect the estimated rover clock error,  since it is identical for all code measurements of constellation $S$.
	Therefore, for the virtual base station, both $dt_b$ and $d_b^S$ are set to zero.
	Second, the multipath and receiver terms $(M^s_{f,b} + \eta^s_{f,b})$ are non-common-mode errors and should not be included.
	Therefore, the desired code measurement to send as an \ac{RTCM} \ac{OSR} message is modeled as
	\begin{equation}\label{eqn:RTCM}
		\left.
		\begin{aligned}
			\rho^s_{f,b} (t_b)  &=R\Big(\BoldP_b, \hat{\BoldP}^s(t^s)\Big)  \\ & ~~ + \hat{E}^s_b(t^s) - c \, (\hat{dt}^s - \delta \hat{t}^s ) + c \, \hat{B}^s_f
			& + \hat{T}^s_b + \hat{I}^s_{f,b}.
		\end{aligned}
	\right\}
	\end{equation}
The server (i.e.,  \ac{VRS}) transmits \ac{RTCM} 3.X messages to the receiver which include the  virtual  base station position $\BoldP_b$ and time $t_b$, and the computed code measurements $\rho^s_{f,b} (t_b)$. 
 The code measurement construction can be decomposed into two parts.
	Section \ref{sec:satpos} will discuss the computation of $R\Big(\BoldP_b, \hat{\BoldP}^s(t^s)\Big)$, which hinges on computation of satellite position $\hat{\BoldP}^s(t^s)$.
	Section \ref{sec:comcorr} will discuss the computation of the common-mode corrections in the second line of Eqn. (\ref{eqn:RTCM}) from the \ac{SSR} data.

%% file: sections/sec_satposest.tex
The purpose of this section is to compute the term $R\Big(\BoldP_b, \hat{\BoldP}^s(t^s)\Big)$ in the first line of Eqn. (\ref{eqn:RTCM})  when  the  base location $\BoldP_b$ and measurement time $t_b$ are given.
This computation must determine the location $\hat{\BoldP}^s(t^s)$ for which the signal received from satellite $s$ transmitted at time $t^s$ would be received at time $t_b$ by a virtual receiver at location $\BoldP_b$.
These computations will neglect the common-mode errors, whose effect is analyzed in Section \ref{sect:CmmnMdEronPos}.

For a signal  pseudo-transmit-time $\hat{t}^s$, 
the \ac{ICD} provides equations 
and the broadcast navigation message provides data
to compute
the satellite clock offset  $\hat{dt}^s(\hat{t}^s)$,
satellite position 
$\hat{\BoldP}^s (\bar{t}^s)$ and satellite velocity $\hat{\BoldV}^s (\bar{t}^s)$, where 
 $\bar{t}^s=({t}_b - t^s_p - \hat{d}t^s)$  is the time-of-signal-transmission corrected by the navigation satellite clock error model from the \ac{ICD}. 
%

\subsection{Algorithm Definition}
The range between the satellite and virtual base antennae can be computed in two ways:
\begin{align}
	R\Big(\BoldP_b, \hat{\BoldP}^s({t}_b - t^s_p - \hat{d}t^s)\Big)	 \mbox{ and }
	c ~ (t^s_p + \hat{d}t^s)
\end{align}
Algorithm \ref{alg:svpos} adjusts the (uncorrected) propagation time $t^s_p$ 
to compute  the satellite position  $\hat{\BoldP}^s(\bar{t}^s)$ 
at the corrected time of transmission $\bar{t}^s$
by minimizing the error between the two range equations:
\begin{align}\label{eqn:F_defn}
	F({ t^s_p}: \BoldP_b, \, {t}_b) 
	&= R\Big(\BoldP_b, \hat{\BoldP}^s({t}_b - t^s_p - \hat{d}t^s)\Big)
	- c ~ (t^s_p + \hat{d}t^s).
\end{align}
The quantity ${t}_g^s = t^s_p + \hat{d}t^s$ is the geometric travel time.
The notation $	F({ t^s_p}: \, \BoldP_b, \, {t}_b) $ means that ${ t^s_p}$ is the argument and $\BoldP_b$ and ${t}_b$ are known parameters.

The optimization is implemented using numeric search. 
Step \ref{alg1:tps} initializes the propagation time to ${t}_p^s = 0.067$ seconds because GNSS satellites are approximately 20,000 km from the Earth surface. 
Step \ref{alg1:tpm} declares a parameter to store the last value of $t^s_p$.
Step \ref{alg1:while} starts the while loop which terminates when $|t^-_p-t^s_p|$ is small.
Step \ref{alg1:t^s} computes the pseudo-transmit time $\hat{t}^s$, 
which is used in Step \ref{alg1:eph} to compute the
satellite 
clock correction $\hat{dt}^s(\hat{t}^s)$, and the
position $\hat{\BoldP}^s(\bar{t}^s)$  and 
velocity $\hat{\BoldV}^s(\bar{t}^s)$.
The \ac{ICD} models \cite{icd2013global,galileo2008galileo,beidou2012beidou} are represented by the function $f(\hat{t}^s,\Boldtheta_e^s)$ where $\Boldtheta_e^s$ are the broadcast ephemeris and clock correction parameters. 
Steps \ref{alg1:r1} - \ref{alg1:F} compute the two ranges and the range error for the current value of ${t}_p^s$.
Step \ref{alg1:partial} computes the partial derivative of the cost with respect to the $t_p^s$ (see App.  \ref{app:partial})
\begin{align}\label{eqn:partial1}
	\frac{\partial F}{\partial {t}_p^s}&=
	 -\left( \1_b^s \, \cdot \, \hat{\BoldV}^s(\bar{t}^s) 	+ c  \right)
	\left(1 - a_1 - 2\,a_2\,(\hat{t}^s-t_{oc})  \right).
\end{align} 
Step \ref{alg1:store} stores the ${t}_p^s$ to $t^-_p$ before updating ${t}_p^s$
Step \ref{alg1:Newton} implements Newton's Zero Finding Algorithm to optimize ${t}_p^s$ to minimize the square of the error defined in Eqn. (\ref{eqn:F_defn}). 
Step \ref{alg1:transt} computes satellite transmission time $\bar{t}^s$  which  will be used as $t^s$ in the \ac{SSR} clock product of Section \ref{sect:orbit_clock}.

For the input receiver time $t_b$ and base station location $\BoldP_b$, this optimization process provides the broadcast clock correction $\hat{dt}^s(\hat{t}^s)$ and the satellite location $\hat{\BoldP}^s(\bar{t}^s)$ at the corrected time of transmission. 
These enable computation of 
$R\Big(\BoldP_b, \hat{\BoldP}^s(\bar{t}^s)\Big)$ 
as required for the \ac{RTCM} \ac{OSR} measurement defined in Eqn. (\ref{eqn:RTCM}).

\begin{algorithm}[tb]
	\caption{Satellite Position Estimation}      
	\label{alg:svpos}               
	\begin{algorithmic}[1]
		\Require $\BoldP_b$, $t_b$.
		\Ensure $\hat{\BoldP}^s(t^s)$, $\hat{V}^s(t^s)$, $\hat{dt}^s$ and $t^s$.
		\State ${t}_p^s = 0.067$; \hfill
		\label{alg1:tps}
		{\color{lgray}// Initialize the propagation time}
		\State $t^-_p = 0$;  \hfill  
		\label{alg1:tpm}
		{\color{lgray}// Initialize previous value to 0}
		\While{$|t^-_p - {t}_p^s|>10^{-11}$}
		\label{alg1:while}
		\State $\hat{t}^s = t_b - {t}_p^s$;  \hfill 
		{\color{lgray}// Pseudo-transmit-time} 
		\label{alg1:t^s}
		\State $[\hat{\BoldP}^s(\bar{t}^s),\hat{\BoldV}^s(\bar{t}^s),\hat{dt}^s(\hat{t}^s)] = f(\hat{t}^s,\Boldtheta_e^s)$; 
		 \hfill  {\color{lgray}// ICD models}
		\label{alg1:eph}
		\State $R_1 = R(\BoldP_b,\hat{\BoldP}^s(\bar{t}^s))$  \hfill 
		{\color{lgray}// Range  based on position}
		\label{alg1:r1}
		\State ${t}_g^s = {t}_p^s + \hat{dt}^s(\hat{t}^s)$ 
		 \hfill {\color{lgray}// See definition in Eqn. (\ref{eqn:F_defn})}
		\State $R_2 =c \cdot {t}_g^s  $  \hfill {\color{lgray}// Range based on geometric travel time}
		\State $F = R_1 - R_2$;	 \hfill 
		{\color{lgray}// Eqn. (\ref{eqn:F_defn}) }
		\label{alg1:F}
		\State $dF = -\left(\frac{\hat{\BoldV}^s(\bar{t}^s)\cdot (\hat{\BoldP}^s-\BoldP_b)}{R_1}+c\right)\,
		 \left(1 - \frac{\partial}{\partial {t}_p^s}\hat{dt}^s(\hat{t}^s)\right)$; \hfill 
		{\color{lgray}// Eqn. (\ref{eqn:partial1})}
		\label{alg1:partial} 
		\State $t^-_p = {t}_p^s$;  \hfill {\color{lgray}// save the previous value}\label{alg1:store} 
		\State ${t}_p^s = {t}_p^s - \frac{F}{dF}$;  \hfill {\color{lgray}// Newton's Zero Finding Algorithm}\label{alg1:Newton} 
		\EndWhile
		\State $\bar{t}^s = t_b - {t}_p^s - \hat{dt}^s(\hat{t}^s)$.   \hfill {\color{lgray}// Transmit time by the satellite clock} \label{alg1:transt} 
	\end{algorithmic}
\end{algorithm}

\subsection{Effect of Common Mode Errors}
\label{sect:CmmnMdEronPos}
At the user receiver, residual measurements are formed for both the base and user receiver measurements. 
For the base measurements, the available information is $t_b$ and $\rho^s_{f,b} (t_b)$.
The user receiver computes the time of satellite transmission as
$$\tilde{t}^s = t_b - \frac{1}{c} \, \rho^s_{f,b} (t_b),$$
where $\tilde{t}_p^s = \frac{1}{c} \, \rho^s_{f,b} (t_b)$ is the measured propagation time. 
Because the measurement of $\rho^s_{f,b} (t_b)$ includes errors, as modeled in Eqn. (\ref{eqn:codeM}),
the measured and actual values of the transmit time are related  according to
$$\tilde{t}^s = \bar{t}^s  - e^s_b,$$
where $e^s_b$ denotes both common-mode and non-common-mode errors scaled to units of time (i.e., $|e^s_b|\le 200$ ns). 
The error $e^s_b$ is neglected in the algorithm of the previous section. 
This section analyzes the effect of this decision on the satellite position and its range from the base. 

The algorithm computes ${\BoldP}^s(\bar{t}^s)$ when it should have computed ${\BoldP}^s(\bar{t}^s-e^s_b)$.
Consider the Taylor series expansion:
\begin{align}
	{\BoldP}^s(\bar{t}^s-e^s_b) &= {\BoldP}^s(\bar{t}^s) - \left.\frac{\partial}{\partial t}{\BoldP}^s({t})\right|_{t=\bar{t}^s} \, e^s_b \\
	&= {\BoldP}^s(\bar{t}^s) - {\BoldV}^s(\bar{t}^s) \, e^s_b
\end{align}
Therefore, because the satellite speed is about 4,000 m/s and $|e^s_b|\le 2\times 10^{-7}$, the effect of neglecting $e^s_b$ on the satellite position calculation is less than one millimeter. 

%% file: sections/subsec_satobtclkcorr.tex
    The \ac{SSR} data streams provide orbit and clock products to correct the satellite position $\hat{P}^s(t^s)$ and clock error $\delta {t}^s(t^s)$ at the transmit time $t^s$. 
    
The orbit correction consists of 
orbit $\delta \BoldO = [\delta O_r,\delta O_a,\delta O_c]^T$ and 
rate $\delta \dot{\BoldO} = [\delta \dot{O}_r,\delta \dot{O}_a,\delta \dot{O}_c]^T$ corrections along with a reference time $t_o$.
These define radial ($\delta O_r$, $\delta \dot{O}_r$), along-track ($\delta O_a$, $\delta \dot{O}_a$) and cross-track ($\delta O_c$, $\delta \dot{O}_c$) components. 
The calculation of the corrected satellite position  $\tilde{\BoldP}^s(t^s)$  from the SSR orbit correction parameters
is presented in Sec. 4.1 of 
\cite{igs2020ssr} (see \cite{rahman2019ecef} for the calculation using the notation of this paper). 
The satellite range ephemeris error $E^s_r$ is computed as 
\begin{equation}
	E^s_b = R(\BoldP_b(t_b), \tilde{\BoldP}^s(t^s)) - R(\BoldP_b(t_b), \hat{\BoldP}^s(t^s)).
\end{equation}

%
%

	The satellite clock correction is transmitted as three polynomial parameters ($C_i^s$, $i = {0,1,2}$) with a corresponding reference time $t_c^s$. 
	The clock correction for satellite $s$ at transmit time $t^s$ is: 
	\begin{equation}\label{eqn:clkcorr}
		\delta C^s(t^s) = C_0^s + C_1^s(t^s - t_c^s) + C_2^s(t^s - t_c^s)~ ~~\text{meters}.
	\end{equation}
	To be consistent with Eqn. (4) in \cite{hadas2015igs}, the correction of satellite clock error $\delta \hat{t}^s$ (i.e., correction  of the broadcast clock error) is 
	\begin{equation}
		\delta \hat{t}^s(t^s) = -\frac{\delta C^s(t^s)}{c}~ ~~\text{seconds}.
	\end{equation}

\black

%% file: sections/subsec_codebias.tex
	Satellite hardware biases have various causes such as analog group delays in the front-end and digital delays.
	They are normally very stable over time (i.e., constant over at least a single day) \cite{sardon1994est}. 
	\ac{GNSS} OS users can apply the \ac{TGD} to roughly correct the hardware bias \cite{ray2005geodetic}. 
	The TGD is computed by the \ac{GNSS} control segment and broadcast in the navigation message for GNSS OS users. 
		
	To attain higher accuracy, \ac{SSR} operations may use either \ac{DCB} or  \ac{OSB} corrections, as defined in
	the SINEX standard \cite{sch2016sinex}. 
	For the \ac{OSB} product,
	let  the symbol  $B_{(\BoldS,T)}$ to denote the correction for observation type $T$ of constellation $\BoldS$ (e.g.,  $T$ could be the C1C observation type with $S$ denoting GPS).
	The $B_{(\BoldS,T)}$ correction products are available in SSR format from sources such as \ac{CAS} GIPP \cite{wang2020gps}.
	For SSR users, the observation 	$O_{(\BoldS,T)}$ is corrected by $B_{(\BoldS,T)}$ as
	\begin{equation}
	O^{'}_{(\BoldS,T)} = O_{(\BoldS,T)} - B_{(\BoldS,T)} \label{eqn:bias}
	\end{equation}	
	to produce the  \ac{OSB}-corrected observation $O^{'}_{(\BoldS,T)}$.
	
	For the OSR measurement in Eqn. (\ref{eqn:RTCM}), 
	\begin{equation}
	\hat{B}^s_f = B_{(\BoldS,T)}. \label{eqn:corrbias}
\end{equation}

%% file: sections/subsec_trop.tex
The troposphere is the lowest layer of the atmosphere, extending to about 60 kilometers above the Earth's surface. 
In this layer, the speed of light is slower than in a vacuum, so the radio signal is delayed. 
The delay is affected by numerous meteorological parameters relevant within the vicinity of the receiver, e.g., temperature, pressure, and relative humidity. 
Numerous tropospheric delay correction models have been introduced \cite{leandro2006unb,li2012new,kazm2017trop}.
These models avoid the need for real-time measurement of the meteorological parameters.

IGGtrop \cite{li2012new} is one such global empirical model.
It computes the slant troposphere delay $T_r^s$ as:
\begin{equation}
T_b^s(E,t) = M(E) ~ T_z(t)
\end{equation}
where 
$E$ is the satellite elevation angle at the rover position;
$$M(E) = 1.001 \left(0.002001 + sin^2( E)\right)^{-\frac{1}{2}}$$ is the mapping function \cite{Penna2001} for converting the \ac{ZTD} to the slant tropospheric delay; and, 
$T_z(t)$ is the \ac{ZTD} (i.e., tropospheric delay in the zenith direction above the receiver). 
IGGtrop computes  $T_z(t)$ as 
\begin{align}\label{eqn:Tzd}
T_z(t) = &a_0 + a_1 \, cos( \gamma \,\, t ) + a_2 \, sin( \gamma \,t )  \\ &
 + a_3 \, cos(2 \,\gamma \,\,t ) + a_4 \, sin(2 \,\gamma \,\,t )  \nonumber
\end{align}
where $\gamma = (2 \pi/365.25)$ and the units of $t$ are real-valued days of the year. 
The coefficients $a_0$, ($a_1$,~$a_2$) and ($a_3$,~$a_4$) model the mean value, the annual variation and the semi-annual variation in \ac{ZTD}, respectively. 
IGGtrop defines these parameters on a 3-D grid in latitude, longitude, and altitude around the globe.
This enables interpolation of the \ac{ZTD} at a receiver position  to achieve 3.86cm root-mean-square error and -0.46cm average bias; however, it requires 666k parameters to implement a global grid.

There are several variants of IGGtrop \cite{li2012new,li2015new,li2018}.
The trade-offs between them related to the required number of parameters and the resulting accuracy. 
The \ac{VNDGNSS} server currently implements the IGGtrop\_SH variant \cite{li2018}.
This approach computes the coefficients $a_0, a_1, a_2, a_3, a_4$ for Eqn. (\ref{eqn:Tzd}) using the  empirical model:
\begin{align} \label{eqn:IGGtrop_exp_func}
	a_0(h) &= exp \bigg( \sum_{i=0}^{m} \alpha_i \, h^i \bigg),\\
	\label{eqn:IGGtrop_poly_func}
	a_j(h) &=  \sum_{i=0}^{5} \beta_{ji} \, h^i  + c, ~ \forall j \in \{1,2,3,4\},
\end{align}
where $h$ is the altitude of the receiver. 
In Eqn.  (\ref{eqn:IGGtrop_exp_func}), the parameter $m$ is a function of the receiver latitude (see Section III in \cite{li2018}). 
The coefficients $\alpha_j$ and $\beta_{ji}$ are defined on a 2-D grid in latitude and longitude.
 These  values were provided by the first author of \cite{li2018}.
The value of $T_z(t)$ is computed at the four nearest points on the 2-D grid and interpolate to compute its value  at the receiver position.

%% file: sections/subsec_iono.tex
	The ionosphere is the ionized zone of the atmosphere containing free electrons and positively charged ions.
	It ranges from about 50 km to 1000 km above the Earth surface. 
	Electromagnetic signals are refracted in the ionosphere (Sec. 6.3 in \cite{teunissen2017springer}). 
	It is a dispersive medium that impacts different frequencies, and the carrier and modulating signals, differently. 
	
	In GNSS OS processes, the dual-frequency user is able to  form ionosphere-free combinations or estimate ionospheric delay using two measurements tracked for two frequencies from one satellite. 
	The single-frequency users apply a prior model (i.e., Klobuchar for GPS and BeiDou \cite{icd2013global,beidou2012beidou,klobuchar1987}, NeQuick for Galileo \cite{galileo2008galileo,gal2020iono})  using ionosphere parameters from the navigation message  to reduce the ionosphere delay by about 50\% \ac{RMS} \cite{yuan2008refi,angrisano2013ass}. 
	Real-time, single-frequency, \ac{PPP} users compensate the ionosphere delay by processing ionosphere products. 
	The VN-DGNSS server includes two possible ionospheric products. 
	Both compute the ionosphere delay for the code measurement on frequency $f$  by
	\begin{equation}
		I_{f,b}^s = \frac{40.3}{f^2}\, STEC ~ 10^{16}.
	\end{equation}
	Each has a different method for computing the \ac{STEC} from a model of the \ac{VTEC}. 
	\ac{VTEC} is computed at the \ac{IPP}, which is computed from the virtual base location $\BoldP_b$ and the computed satellite location $\hat{\BoldP}^s(\bar{t}^s)$.

	{\bf SSR VTEC-SH: }
	The agencies listed in Table \ref{tab:vtec} provide real-time \ac{VTEC} \ac{SSR} products that have global applicability. 	
	In the \ac{RTCM} Version 3 standard \cite{rtcm2014proposal}, the real-time \ac{VTEC} \ac{SSR} message provides the parameters for a \ac{SH} expansion with
	degree of the expansion $N$, 
	order of the expansion $M$, and 
	coefficients  $C_{n,m}$ and  $S_{n,m}$ for $n=0,\ldots,N$ and $m=0,\ldots,M$.
	For a single-layer model, the \ac{VTEC} model at the \ac{IPP} is  \cite{li2020igs,igs2020ssr}
	\begin{align}
		VTEC(\phi_{PP},\lambda_{PP}) &= \sum^{N}_{n=0} \sum^{min(n,M)}_{m=0} P_{n,m} (\sin(\phi_{PP})) \\
		&\cdot (C_{n,m}\,\cos(m \, \lambda_{S})+S_{n,m}\,\sin(m \, \lambda_{S}))
	\end{align}
	where 
	$P_{n,m}()$ is the normalized associated Legendre function, 
	$(\phi_{PP},\lambda_{PP})$ are the geocentric latitude and longitude of the \ac{IPP}, and 
	$\lambda_{S}$  is computed by 
	\begin{equation}
		\lambda_{S} = (\lambda_{PP}+(t-50400)*\pi/43200)\,modulo\,(2\pi)
	\end{equation}	
	where $t$ is the \ac{SSR} epoch time of computation,  modulo 86400 seconds.

	The \ac{STEC} is  computed from \ac{VTEC} using
	\begin{equation}
		STEC = \frac{VTEC}{sin(E+\phi_{PP})}
	\end{equation}
	where $E$ is the elevation angle of the satellite at the rover position.

\begin{table}[h]
	\centering
	\caption{Public real-time global ionosphere VTEC modeling services providing RTCM VTEC message. Each has NTRIP caster ID: `products.igs-ip.net:2101'.} 
	\begin{tabular}{c|c|c|c}
		\hline
		Agency   & SH degree & Mount point \T &   References        \\ \hline
		CAS   & 15  &    SSRA00CAS0 \T&  \cite{li2015shpts}  \\ \hline
		CNES   & 12  &   SSRA00CNE0 \T&  \cite{cne2015vtec}  \\ \hline
	\end{tabular}
	\label{tab:vtec}
\end{table}

	{\bf US-TEC: }
	For North America users, the \ac{NOAA} provides the real-time \ac{USTEC} product for public usage \cite{fuller2005ustec}. 
	\ac{USTEC} provides \ac{VTEC} values in a uniform grid of point locations with 1 degree resolution in latitude and longitude. 
	The official website \cite{ustecweb} states that   the normal update interval is 15 minutes, typically with 28 minutes latency\footnote{During the period from late 2020 through the middle of 2021, the latency was over 24 hours; therefore, although US-TEC is included in the software distribution, it was not used\label{ftnt:noUSTEC} in the demonstration experiments described in Section \ref{sec:test}. }.

	The \ac{VTEC} at the \ac{IPP} (within the geographical extent of the grid points) is computed as a linear function of the values at grid points.
	The \ac{VNDGNSS} server uses distance weighted spatial interpolation \cite{el1993faa,prol2017comparative}:
	\begin{equation}\label{eqn:tecgrd}
		VTEC(IPP) = \Boldw(IPP)~\BoldI^G,
	\end{equation}
   where $\Boldw(IPP)$ is a vector of weights. 
   Both $\Boldw(IPP)$ and $\BoldI^G$ are vectors with $K$ components. 
   The $k$-th element $w_k$ of the weighting vector determines the amount that a \ac{VTEC} value $[\BoldI^G]_k$ at the $k$-th grid point $IPP_k$ contributes to the value of \ac{TEC} at the desired \ac{IPP}. 
   
   The vector $\Boldw(IPP) = [w_1,...,w_K]$  satisfies the following constraints:
	\begin{enumerate}
		\item $\sum^{K}_{i=1}w_k = 1$, so that Eqn. (\ref{eqn:tecgrd}) interpolates the value for VTEC from the grid points.
		\item the $k$-th element $w_k$ should decrease smoothly as $d_k = \|IPP-IPP_k\|$ increases. 
		This results in a smooth interpolation between the grid point values.
	\end{enumerate}
		The approach herein uses at most four nonzero weights. If $IPP = IPP_k$ for some k, then $w_k = 1$ and all remaining values of $\Boldw(IPP)$ are zero. 
		Otherwise, the IPP is between four grid points. 
		In this case, we define $a\in\Re^K$ such that $a_k = 1/d_k$ for those four corner grid points, noting that $a_k$ is finite because $d_k \neq 0$.
		All the remaining elements of $\Bolda$ are zero. 
		Then, the components of $\Boldw(IPP)$ are defined as
	$
		w_k = \frac{a_k}{\sum_{i=1}^K a_i}.
	$

    Spatial interpolation provides the \ac{VTEC} at the \ac{IPP}. 
    The \ac{STEC} is computed as \cite{prol2017comparative}:
	\begin{equation}
		STEC = F(E)  \,   VTEC(IPP).
	\end{equation}
    where $F(E)$ is the ionospheric obliquity factor defined as	
    \begin{equation}
    	F(E) = \frac{1}{\sqrt{1 - \left[\frac{r_e \, \cos(E)}{r_e + h_m}\right]^2}},
    \end{equation}
	where $r_e$ is the average radius of the earth, and $h_m$ is the height of the maximum electron density (assumed herein to be 350 km).

%% file: sections/sec_newdesign.tex
\section{VN-DGNSS Server-Client System design} \label{sec:NewDes}
This section describes the software implementation for a publicly available, open-source,
 client/server \ac{VNDGNSS} implementation  with global coverage.
The server receives real-time information in \ac{SSR} format and provides each client with virtual base station \ac{RTCM} \ac{OSR} formatted messages applicable to their local vicinity for BeiDou B1, GALILEO E1, and GPS L1. 
This approach eliminates the need for the client to have physical access to a local reference station.

\begin{figure}[tb]
	\centering    
	\includegraphics[width=\linewidth]{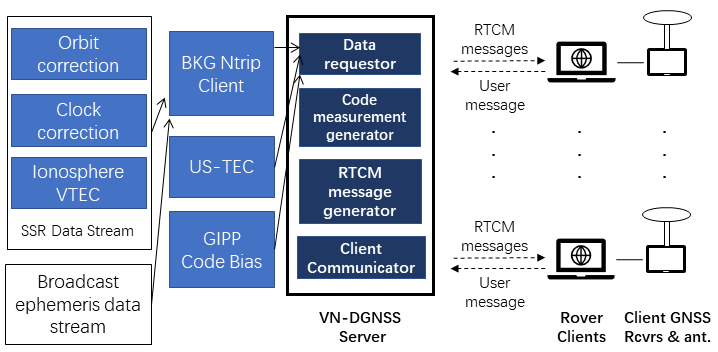}
	\caption{VN-DGNSS server-client architecture}\label{fig:VND}
\end{figure}

\subsection{System Architecture}
Fig. \ref{fig:VND} displays the \ac{VNDGNSS} client/server framework. 
The \ac{VNDGNSS} server includes four fundamental components. 
\begin{itemize}
	\item The {\em data requestor} establishes communications with online sources to maintain real-time \ac{SSR} data.
	The \ac{BNC} \cite{weber2016bkg} provides 
	broadcast ephemeris for each satellite system,  
	Multi-GNSS \ac{SSR} orbit and clock corrections, and SSR VTEC information (see Table \ref{tab:vtec}).
	\ac{USTEC} products are included as an option for US clients.
	\ac{CAS} GIPP provides code bias products \cite{wang2020gps}.   
	Because the IGGtrop\_SH parameters are static, they are loaded at run-time and do not require real-time communications. 
	\item The {\em code measurement generator} constructs the code measurements, for the virtual base position $\BoldP_b$ requested by the client,  using  Eqn. (\ref{eqn:RTCM})  for GPS L1, Galileo E1 and Beidou B1.
	The range is computed as described in Section \ref{sec:satpos}.
	The correction terms in the second line of Eqn. (\ref{eqn:RTCM}) are computed using the models discussed in Section \ref{sec:comcorr}.

	\item The {\em RTCM message generator} creates RTCM message type 1004 and MSM4 
	message types: 1074 for GPS, 1094 for Galileo, 1124 for BeiDou.
	The RTCM 1004 message communicates the virtual reference station position $\BoldP_b$.
	The RTCM MSM4 message includes the code measurements constructed by the code measurement generator at integer epoch times $t_b$ (i.e., 1.0 Hertz), along with the observation type and \ac{SNR}\footnote{In the current implementation, the server computes the SNR as 
	$floor(\frac{A ~ E}{\pi}) + \underline{S}$, where $E$ is the satellite elevation in radians, $\underline{S}$ is the minimum SNR, and $A$ is a coefficient that determines the maximum SNR.   }.

	\item The {\em client communicator} receives requests from, then establishes and maintains TCP connections with clients. 
	In addition to its internet address, the client sends its (desired) virtual reference station position $\BoldP_b$, and the GNSS constellations and the observation types for which the client desires RTCM messages.
	Clients that are moving can update their desired virtual reference station position as necessary to keep it within a reasonable distance (e.g., 20 km) of their  location.  
\end{itemize}
 Some portions of the \ac{VNDGNSS} (e.g.,  code measurement generator, RTCM message generator) are modifications of  open-source functions in RTKLIB \cite{rtklibgit,takasu2009development} and \ac{BNC} \cite{weber2016bkg}.

\subsection{Implementation Strategies}
Details and choices required for the \ac{VNDGNSS} server implementation are described below.
\begin{enumerate}
\item Table \ref{tab:vtec} list the  two agencies that currently stream SSR Multi-GNSS corrections and global VTEC SH parameters  using \ac{RTCM} messages through NTRIP. 
Both are available through the server.

\item The real-time orbit and clock corrections are applicable to a specific \ac{IOD} ephemeris. 
GPS and Galileo ephemeris provides the IOD number for each satellite.
The Beidou IOD depends on the agency providing the data. 
The two  agencies in Table \ref{tab:vtec} both share the same methodology: $$BeiDou~IOD = ((int) toe/720)~modulo~240,$$ where $toe$ is the time-of-ephemeris in the BeiDou navigation message \cite{pppwizard}. 

\item The multi-GNSS \ac{OSB} product provided by \ac{CAS} \cite{wang2020gps} is  referred to as GIPP.
The GIPP code bias data can be downloaded from \url{ftp://ftp.gipp.org.cn/product/dcb/daybias/}.
They are updated with a period of one day update and latency of 1-5 days.
The two agencies listed in Table \ref{tab:vtec} also provide code biases through NTRIP; however, they do not provide bias estimates for all observation types required for the experiments.

\item 
The SSR data has two types: \ac{APC} and \ac{COM}.
The broadcast ephemeris is referred to the satellite's \ac{APC}; 
therefore, the selected \ac{SSR} data stream for orbit corrections must also reference to the \ac{APC}, not  \ac{COM}.

\end{enumerate}

%% file: sections/sec_expresults.tex
The \ac{VNDGNSS} evaluation in this section includes data from both stationary and moving platforms.
This section describes the 
receiver hardware and configurations, 
virtual network data sources,
metrics,  and 
experimental results. 

\subsection{Experiment Schedule}
Results from three experiments are included herein\footnote{Additional information about the results of the experiments are described in the document  `test\_data/Experiment\_Support\_Doc /support\_doc.pdf'  at the GitHub repository defined in Footnote \ref{ftnt:GitHub}). \label{ftnt:expRslts}} from one stationary test and two moving tests. 

The {\em stationary} test used a dual-band antenna (Trimble ACCG5ANT\_2AT1) located on the roof of Winston Chung Hall at \ac{UCR}.
The ground truth location of this antenna are known to centimeter accuracy from prior \ac{OPUS} survey results.
VN-GNSS position estimates from the receiver  are compared against that {\em ground truth} location.

For the {\em moving platform} experiments, a u-blox antenna is mounted on the roof of a car.
Using a signal splitter, that antenna is connected to 4 receivers for each experiment. 
One u-blox M8P receiver is used to determine an independent {\em ground truth} trajectory for the time-varying antenna position.
This receiver was communicating with a nearby base station and performing carrier-phase, integer-fixed \ac{RTK} to achieve centimeter accuracy.
The configurations of the remaining 3 receivers for each moving test are discussed in Section \ref{sect:moving test} using the experiment labels from Table 5.
The vehicle was driven in the urban area near \ac{UCR} that had a relatively clear view of the sky. 
The duration of each data accumulation experiment was about 1 hr.

\subsection{GNSS Hardware and Configuration Descriptions}
\label{sect:moving_configs}
The experiments used 5 low-cost receivers, each compatible with \ac{RTCM} \ac{OSR} messages. 
The text below describes and Table \ref{tab:term} summarizes the 6 experimental configurations that are compared. 
 The acronym VN is used to indicate  \ac{VNDGNSS}. \black

 Two u-blox M8P receivers were used. 
	Each M8P  is capable of tracking GPS, GLONASS, and Beidou satellites. 
	\ac{VNDGNSS} does not currently incorporate GLONASS corrections, so GLONASS was disabled.
	With the current firmware (version 3.01 HPG 1.40), 
	for Beidou, the M8P will use the \ac{RTCM} corrections only if available for both code and phase;
	 therefore, because \ac{VNDGNSS} does not currently provide phase measurements, the BeiDou measurements were disabled.
	Therefore, only GPS measurements were used with the M8P.
	\begin{description}
		\item[SF GPS SPS: ] An M8P  was configured for \ac{SPS}\footnote{For GPS, the open service is usually stated as \ac{SPS} \cite{team2014global}, which we use in the experiment label. 
		Columns 2-4 of Table  \ref{tab:term} use the acronym \ac{OS} instead of \ac{SPS} for consistency in those columns.
		} without any external correction sources, just the standard tropospheric and ionospheric models.
		\item[SF GPS VN: ] An M8P connects to the \ac{VNDGNSS} client to receive real-time RTCM OSR corrections.
	\end{description}

 Three u-blox ZED-F9P receivers were used. 
Each F9P is a dual-frequency receiver compatible with Multi-GNSS.	
GPS, Galileo, and BeiDou are used herein. 
	It used \ac{SF} operation when connected to a single band antenna or \ac{DF} operation when connected to a dual band antenna.
	\begin{description}
		\item[SF GNSS OS: ] An F9P was configured to operate in \ac{SF} \ac{OS} mode (i.e., \ac{SBAS} disabled and no \ac{VNDGNSS}). 
		\item[DF GNSS OS: ] An F9P is configured to operate in \ac{DF} \ac{OS} mode (i.e., \ac{SBAS} disabled and no \ac{VNDGNSS}).
		\item[SF GNSS VN: ] An F9P is configured to use \ac{SF} \ac{RTCM} \ac{OSR} messages through the \ac{VNDGNSS} client.
		When the client connects to the server (typically within about one second), the F9P switches to \ac{DGNSS} mode.
		\item[F9P SBAS:   ] An F9P  was configured with \ac{SBAS}  enabled. 
		Because the experiments occurred in California, they were within the \ac{WAAS} coverage area, allowing differential operation for GPS.	
		The F9P typically requires a few minutes to start tracking a \ac{WAAS} augmentation satellite, at which time it switches to \ac{DGPS} mode and process only the L1 measurements. The GALILEO and BeiDou satellite measurements were tracked and used by the receivers in OS mode.
	\end{description}

\black
All receivers report their computed position every 1 second using the manufacturer algorithms.  The elevation angle cutoff for all the receivers was 15$^\circ$.

\begin{table}[bt]
	\centering
	\caption{Descriptions of experimental configurations.} 
	\begin{tabular}{@{\hskip3pt}c@{\hskip3pt}|c|c|c|c|c}
		\hline
		\multirow{2}{*}{\begin{tabular}[c]{@{}c@{}}Experiment\\ label\end{tabular}} & \multicolumn{3}{c|}{GNSS Systems Used \T } & \multirow{2}{*}{SBAS} & \multirow{2}{*}{Receiver} \\ \cline{2-4}
						& GPS\T   & GALILEO & BeiDou  &                       &                           \\ \hline
		SF GPS SPS      & OS     & None    & Disable    & None                  & M8P\T                       \\ \hline
		SF GPS VN       & VN      & None    & Disable    & None                  & M8P\T                        \\ \hline
		SF GNSS OS      & OS      & OS      & OS      & Off               & ZED-F9P\T                    \\ \hline
		DF GNSS OS      & OS      & OS      & OS      & Off               & ZED-F9P\T                    \\ \hline
		SF GNSS VN      & VN     & VN    & VN     & Off               & ZED-F9P\T                    \\ \hline
		F9P SBAS        & SBAS    & OS      & OS      & WAAS              & ZED-F9P\T                    \\ \hline
	\end{tabular}
	\label{tab:term}
\end{table}
\black

\subsection{Data Choices}
\label{sect:data_choices}
Because the accuracy assessments for the real-time \ac{VTEC} products \cite{li2020igs} and for the orbit and clock products \cite{wang2018assessment} indicate that \ac{CNES} has slightly better performance than \ac{CAS}, we select CNES as the data source in these experiments.
The reason why US-TEC was not used is stated in Footnote \ref{ftnt:noUSTEC}.

Table \ref{tab:accobt} summarizes the predicted error statistics based on the following articles:
\ac{CNES} real-time orbit and clock products \cite{wang2018assessment};
GIPP OSB  \cite{wang2018code,wang2020gps}; 
 IGGtrop\_SH  \cite{li2018}; \black
 \ac{CNES} \ac{VTEC}   \cite{li2020igs};
 along with the non-common mode errors.
The VTEC errors are frequency dependent;
therefore, they are slightly different for each satellite systems.
The IGGtrop\_SH \ac{RMS}  error is the same for all satellite systems because the model is  for properties of the environment that are independent of frequency and of the satellite system.
Based on the information in this table, multipath is the dominant remaining error source.

\begin{table}[b]
	\centering
	\caption{ \ac{VNDGNSS} ranging errors. 
	} 
	\begin{tabular}{@{\hskip4pt}r@{\hskip4pt}|c|c|c}
		\hline
		Error 				& GPS (L1)  & Galileo (E1) & BeiDou (B1)\T \\ \hline
		CNES Orbit 3D (cm) RMS	& 2.88 & 4.42    & 14.54 \T \\
		CNES Clock (ns) RMS		& 0.45 & 0.39   & 3.00  \\
		CAS GIPP (ns) RMS		& 0.06 & 0.15   & 1.0  \\
		IGGtrop\_SH (cm) RMS		& 3.86 & 3.86 & 3.86 \\
	    CNES VTEC (cm) RMS 		& 55.21 & 55.21 & 56.22\\
		Code Multipath (m)		& $<$ 2-3   &  $<$ 2-3  &  $<$ 2-3 \\
		Noise	(mm)			& $<$ 1  & $<$ 1   &  $<$ 1  \\ \hline
	\end{tabular}
	\label{tab:accobt}
\end{table}

\subsection{Metrics}
\label{sect:metrics}
The  metrics for analyzing position error will include: 
\begin{itemize}
	\item ECEF 3-D  error norm : $\|\delta\BoldP^e\|$ with $\delta\BoldP^e = \hat{\BoldP}^e-\BoldP^e$;
	\item Horizontal  error: $HE = \sqrt{ (\delta\BoldP^g_N)^2 + (\delta \BoldP^g_E)^2}$;
	\item Vertical  error: $VE = |\delta \BoldP^g_D|$.
	
\end{itemize}
The notation in these expressions is as follows. 
The symbol $\BoldP$ denotes the ground truth position; and,
$\hat{\BoldP}$ denotes the position estimated by the receiver.
The superscript denotes the frame-of-reference: $e$ for \ac{ECEF} and $g$ for local tangent frame. 
Vectors are transformed between frames using $\delta \BoldP^g = \BoldR_{e}^g \, \delta\BoldP^e$,
where the symbol
$\BoldR_e^g$ represents the rotation matrix from the \ac{ECEF} frame to the local \ac{NED} frame (see Eqn. (2.24) in \cite{cai2011unmanned}).
In the local tangent frame, the coordinates of $\BoldP^g$ and $\delta\BoldP^g$ are indicated by $N$, $E$, and $D$ subscripts.

Each of these metrics will be analyzed with respect to the following standards:
$Pr\{HE\le 1 \hbox{ m}\}$,        $Pr\{VE\le 3 \hbox{ m}\}$, and
$Pr\{\|\delta\BoldP^e\|\le 3 \hbox{ m}\}$, where
$Pr\{x\le \tau\}$    denotes the probability that $x$ is less than threshold $\tau$.
Note that the SAE standards \cite{sae2016} require that
$Pr\{HE\le 1.5 \hbox{ m}\}\ge 68\%$ and        $Pr\{VE\le 3 \hbox{ m}\}\ge 68\%$.

\subsection{Stationary Test Results}\label{sect:stationary test}

The experiment uses the receiver configurations labeled:
 SF GPS SPS, SF GPS VN,
 DF GNSS OS,    SF GNSS VN, and F9P SBAS.

\begin{figure}[tb]
	\includegraphics[width=\linewidth]{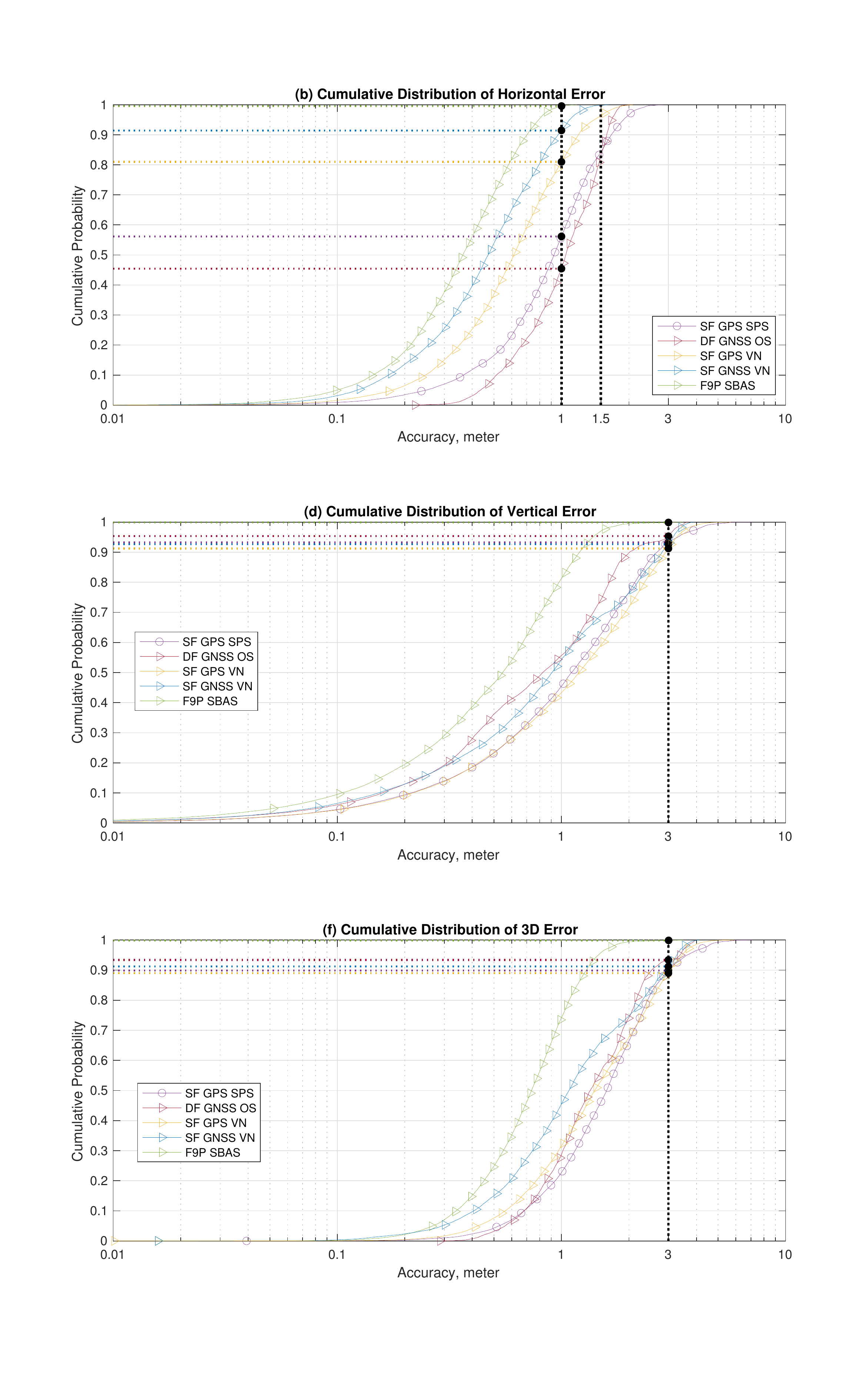}
	\caption{Stationary experimental results. 
		Subplots (a), (b), and (c) show the cumulative error distributions for the horizontal, vertical, and 3D position error metrics:
		 blue for SF GPS SPS, 
		 orange for DF GNSS OS,
		 green for SF GPS VN,
		 purple for SF GNSS VN, and
		 yellow for F9P SBAS.}
	\label{fig:test_sta}
\end{figure}

The test period is about 12 hours,
with approximately  $4.3\times 10^3$ measurement epochs.
Fig. \ref{fig:test_sta} displays the cumulative error probability distributions for the error metrics defined in Section \ref{sect:metrics}. 
The temporal error plots are also available (see Footnote \ref{ftnt:expRslts}). 
The statistics are summarized in Table \ref{tab:sta}.
Table \ref{tab:sta} and Fig. \ref{fig:test_sta} allow the following conclusions for the stationary data:
\begin{itemize}
	\item SBAS performs the best with at least a 4-8\% advantage in all the metrics.
	Note that this service is only available to users with SBAS capable receivers in the SBAS coverage area 
	(e.g., WAAS for GPS in North America, EGNOS for Galileo in Europe, and BDSBAS for BeiDou in China). 
	
	\item SF GPS VN and SF GNSS VN both yield significantly better horizontal accuracy than SF GPS SPS or DF GNSS OS. 
	The  \ac{VNDGNSS} approach improves the horizontal sub-meter level accuracy by 24.84\% relative to SF GPS SPS and 46.03\% relative to DF GNSS OS.  
	
	\item
	All the configurations have vertical performance surpassing the SAE specification.
	F9P SBAS performs the best at all accuracy levels. 
	DF GNSS OS performs second best.
	Vertical errors are significantly affected by atmospheric delays, which are best accommodated by 
	two-frequency processing or, for users in N. America,  the high-precision atmospheric model in WAAS.
	 Fig. \ref{fig:test_sta}(b) shows  SF GNSS VN performance approaches that of DF GNSS OS at smaller error levels. 
	\blue
\end{itemize}

\begin{table}[bt]
	\centering
	\caption{ Stationary antenna error probability summary.} 
	\begin{tabular}{c|c|c|c}
		\hline
		Stationary   &   $Pr\{HE\le 1 \hbox{ m}\}$      &   $Pr\{VE\le 3 \hbox{ m}\}$\T \B  
		&   $Pr\{\|\delta\BoldP^e\|\le 3 \hbox{ m}\}$     \\ \hline
		SF GPS SPS       & 56.16\% 							&  93.27\% 					     & 89.89\% \T \\ \hline
		SF GPS VN      & 81.00\%    					   & 91.23\%  					   & 89.01\%  \T \\ \hline
		DF GNSS OS          & 45.43\%  						& 95.30\%   				   & 93.37\% \T \\ \hline
		SF GNSS VN  & 91.46\%  				    & 92.60\%   				      & 91.19\% \T \\ \hline
		F9P SBAS           & 99.59\%  						  & 99.88\%   				      & 99.87\% \T \\ \hline
	\end{tabular}
	\label{tab:sta}
\end{table}

\subsection{Moving Test Results} \label{sect:moving test}
Two experiments were performed over distinct time intervals  using two u-blox antennae as follows:
\begin{enumerate}
	\item The SF GNSS OS, SF GNSS VN, and F9P SBAS receiver configurations were performed using a single-band u-blox antenna.
	\item The DF GNSS OS, SF GNSS VN, and F9P SBAS receiver configurations were performed using a dual-band u-blox antenna.
\end{enumerate}
\black

\begin{figure*}[htbp]
	\includegraphics[width=\textwidth]{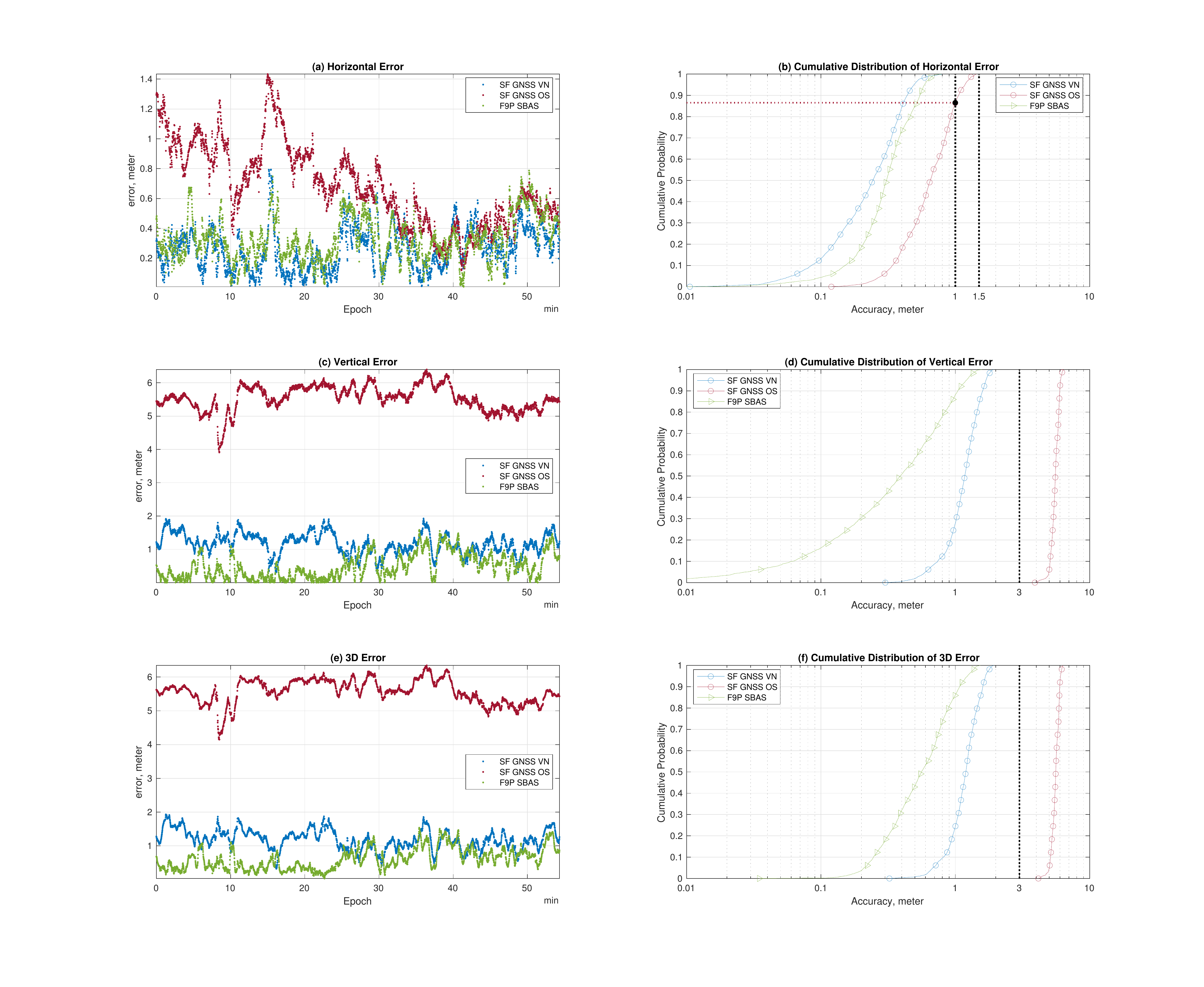}
	\caption{Experimental results for a moving platform test using a single-band antenna. 
		Subplots (a), (c), and (e) show the horizontal, vertical, and 3D position error plotted versus time. 
		The $x$-axis is the epoch number. 
		Subplots (b), (d), and (f) show the cumulative error distribution of the horizontal, vertical, and 3D position error metrics:
		blue for SF GNSS VN, 
		red for SF GNSS OS, and
		green for F9P SBAS.
		}
	\label{fig:test_movSF}
\end{figure*}

\black
For the test using a single-band antenna experiment, the summary in Table \ref{tab:mov_sin} shows that the SF GNSS VN and F9P SBAS configurations both surpass the SAE specifications for both horizontal and vertical accuracy. 
Both also surpass the 1m horizontal accuracy at 95\% specification stated in Table 1 of \cite{NigelBarth2021} for Lane-Level applications.
Both SF GNSS VN and F9P SBAS improve horizontal sub-meter level accuracy by 13.47\% relative to SF GNSS OS.
Fig. \ref{fig:test_movSF}(b) shows that SF GNSS VN has better horizontal performance than SF GNSS OS and F9P SBAS at all accuracy levels. 
Fig. \ref{fig:test_movSF}(d) shows that F9P SBAS achieved better vertical accuracy than SF GNSS VN, with the SF GNSS OS vertical performance failing to achieve the SAE specification.

\begin{figure*}[tbp]
	\includegraphics[width=\textwidth]{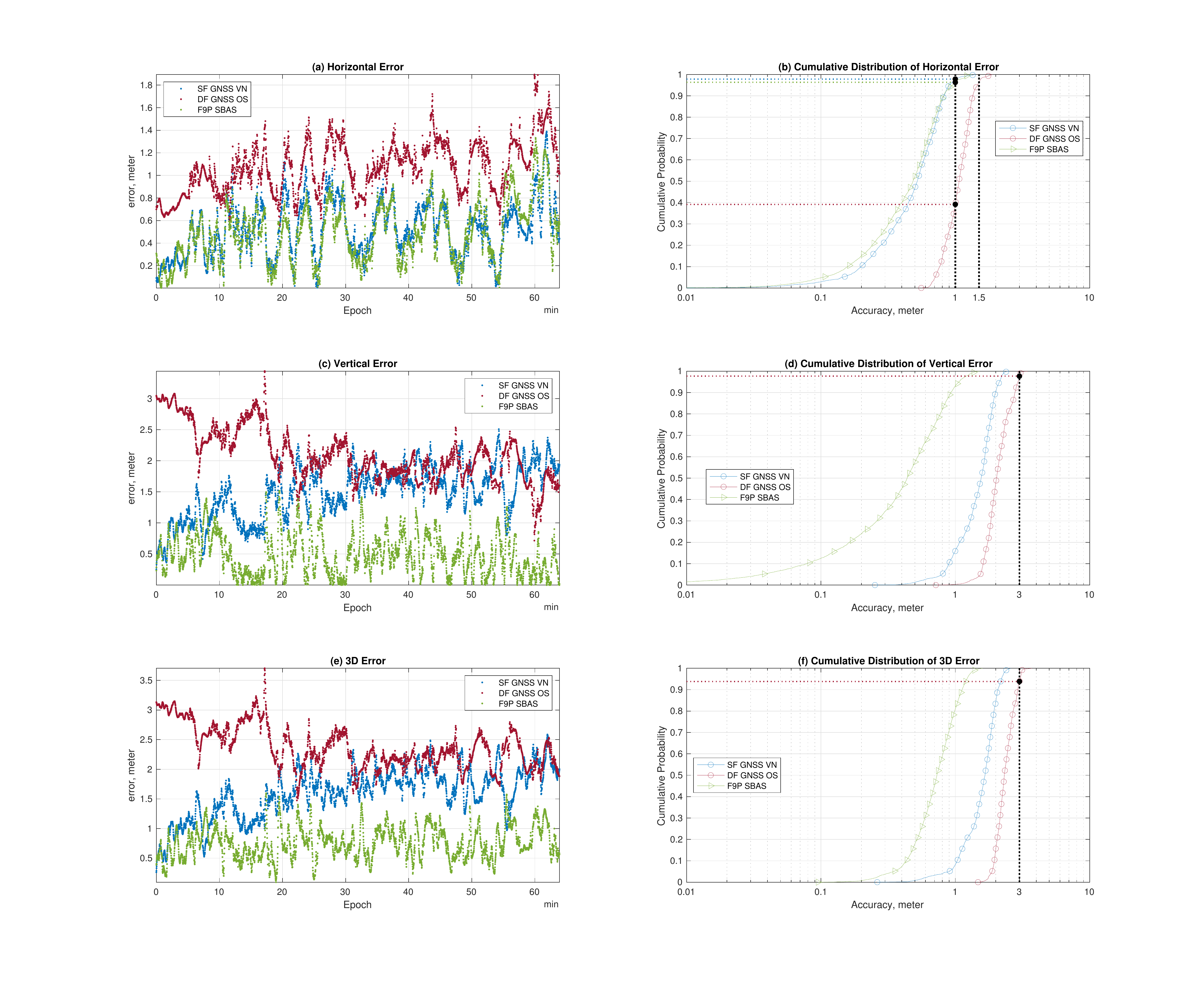}
	\caption{Experimental results for a moving platform test using a dual-band antenna. 
		Subplots (a), (c), and (e) show the horizontal, vertical, and 3D position error plotted versus time. 
		The $x$-axis is the epoch number. 
		Subplots (b), (d), and (f) show the cumulative error distribution of the horizontal, vertical, and 3D metrics:
		blue for SF GNSS VN; 
		red for DF GNSS OS; and
		green for F9P SBAS.}
	\label{fig:test_movDF}
\end{figure*}

For the test using a dual-band antenna experiment, Fig. \ref{fig:test_movDF} and Table \ref{tab:mov_dual} show that 
SF GNSS VN and F9P SBAS achieve comparable performance which is significantly better than DF GNSS OS by greater than 55\%.
Fig. \ref{fig:test_movDF}(d) shows that all three configurations surpass the SAE vertical accuracy specification. 
F9P SBAS  achieves the best vertical accuracy, 
while SF GNSS VN surpasses DF GNSS OS.

\begin{table}[bt]
	\centering
	\caption{SF, moving platform, error  summary. } 
	\begin{tabular}{@{\hskip4pt}c@{\hskip4pt}|@{\hskip4pt}c@{\hskip4pt}|@{\hskip4pt}c@{\hskip4pt}|@{\hskip4pt}c@{\hskip4pt}}
		\hline
		Moving SF  &   $Pr\{HE\le 1 \hbox{ m}\}$      &   $Pr\{VE\le 3 \hbox{ m}\}$ \T  \B
		&   $Pr\{\|\delta\BoldP^e\|\le 3 \hbox{ m}\}$     \\ \hline
		SF GNSS OS   \T     & 86.53\% 							&  0.00\% 					     &0.00\% \\ \hline
		SF GNSS VN   \T     & 100.00\%    					   & 100.00\%  					   & 100.00\% \\ \hline
		F9P SBAS     \T        & 100.00\%  						  & 100.00\%   				      & 100.00\%\\ \hline
	\end{tabular}
	\label{tab:mov_sin}
\end{table}

\begin{table}[bt]
	\centering
	\caption{DF, moving platform, error  summary.} 
	\begin{tabular}{@{\hskip4pt}c@{\hskip4pt}|@{\hskip4pt}c@{\hskip4pt}|@{\hskip4pt}c@{\hskip4pt}|@{\hskip4pt}c@{\hskip4pt}}
		\hline
		Moving DF   &   $Pr\{HE\le 1 \hbox{ m}\}$      &   $Pr\{VE\le 3 \hbox{ m}\}$ \T\B  
		&   $Pr\{\|\delta\BoldP^e\|\le 3 \hbox{ m}\}$     \\ \hline		
		DF GNSS OS   \T   & 39.14\% 							&  93.82\% 					     & 97.65\% \\ \hline
		SF GNSS VN   \T  & 97.86\%    					   & 100.00\%  					   & 100.00\% \\ \hline
		F9P SBAS    \T    & 96.30\%  						  & 100.00\%   				      & 100.00\%\\ \hline
	\end{tabular}
	\label{tab:mov_dual}
\end{table}

%% file: sections/sec_conclusion.tex
\section{Discussion}
\label{sect:disc3}

Table \ref{tab:accobt} predicts that multipath and ionospheric delay error will be the two largest remaining ranging errors.
These predictions are supported by the reported experiments.

The effects of multipath are distinct for stationary and moving receivers.
Multipath is caused by nearby reflective surfaces. 
For stationary antennae and reflectors, multipath can be correlated over long periods of time. 
For a moving platform, the receiver moves relative to the reflective surface; therefore, the multipath errors change much more rapidly (i.e., less correlated across time),
which makes them easier for the receiver to filter out during the vehicle state estimation process.
This is one explanation for the improved horizontal performance for the moving platform relative to the stationary platform. 

Because the sign of the L1/E1/B1 code measurement ionospheric delay term is always positive, and 
the signs of the coefficients are always positive in the measurement observation matrix for the clock and Down-position components, the atmospheric delay most strongly contributes to error in the clock and Down estimates. 
Figures \ref{fig:test_sta}(d), \ref{fig:test_movSF}(d), and \ref{fig:test_movDF}(d) clearly show that SBAS (i.e., WAAS) achieves the best vertical accuracy.
The WAAS performance analysis report \cite{waas2020report}  does not provide explicit assessment for WAAS ionosphere correction, but list the position horizontal and vertical errors as less than 1m at 95\% for most sites in the continental United States.
These performance levels imply that the WAAS ionosphere grid based model, which was originally designed for avionics applications within North America, achieves more accurate ionospheric delay predictions in southern California than does the CNES VTEC spherical harmonic model, which was designed for a global scale.

\section{Conclusions and Future Research}
\label{sect:concl}
This paper presents, demonstrates, and evaluates a  VN-DGNSS approach that is built on a client-server architecture.
Usage of the server is free.
The source code is available via open-source at the GitHub repository described in footnote 1.
The VN-DGNSS server transmits RTCM OSR messages computed from real-time SSR data sources.
RTCM SC-104 is a standard that defines the data structure for OSR differential correction information supported universally by almost all GNSS receivers. 

The single-frequency  code-measurement-only  experiments presented in Sections \ref{sect:stationary test} and \ref{sect:moving test} demonstrate that the VN-DGNSS succeeds in improving positioning accuracy relative to GNSS OS, 
without the user needing a physical base station near the user.
The position estimation performance is comparable to that achieved using WAAS, but is available globally.
This is valuable to users of RTCM enabled receivers that are not SBAS enable or that are operating in a region of the world not covered by an SBAS implementation.

There are several directions that could be considered for future research.
This paper and the currently implemented system focus on single-frequency code measurements for  navigation systems using \ac{CDMA}. 
The extension to GLONASS, which uses \ac{FDMA}, is more challenging.
As discussed in Sect. 21.1.1 in \cite{teunissen2017springer} and footnote \ref{ftnt:FDMA},
there will be frequency-dependent biases between FDMA receiver's channels, that are different between receivers and are unknown.
The \ac{VNDGNSS} approach could be extended to correct for GLONASS
satellite orbit and clock errors, 
satellite code bias, 
satellite frequency dependent biases, and 
atmospheric errors. However, this would leave the receiver \ac{FDMA} ICBs as significant sources of error.
\black
Extensions to multi-frequency and phase measurements are both interesting. 
The approach described herein can be extended to multiple frequencies per system; however, the scope of the funded project was limited to formulation and demonstration of the single frequency approach.
For phase-measurement-based positioning, namely \black
\ac{RTK}, which is supported by many GNSS receivers. 
Sec. \ref{sect:disc1} discusses the basic reasons why phase measurements are not currently supported by the VN-DGNSS for \ac{RTK}.
\ac{PPP-AR} using \ac{SSR} products is a very active area of research \cite{geng2019pride,laurichesse2010real,li2018multi} achieving centimeter to decimeter accuracy. 
A VN-DGNSS approach that packages the SSR products as OSR RTCM messages should be able to achieve these same accuracies using PPP-like methods. 
Research into such approaches is ongoing.

%% file: biography/Biography.tex
\begin{IEEEbiography}[{\includegraphics[width=1in,height=1.25in,clip,keepaspectratio]{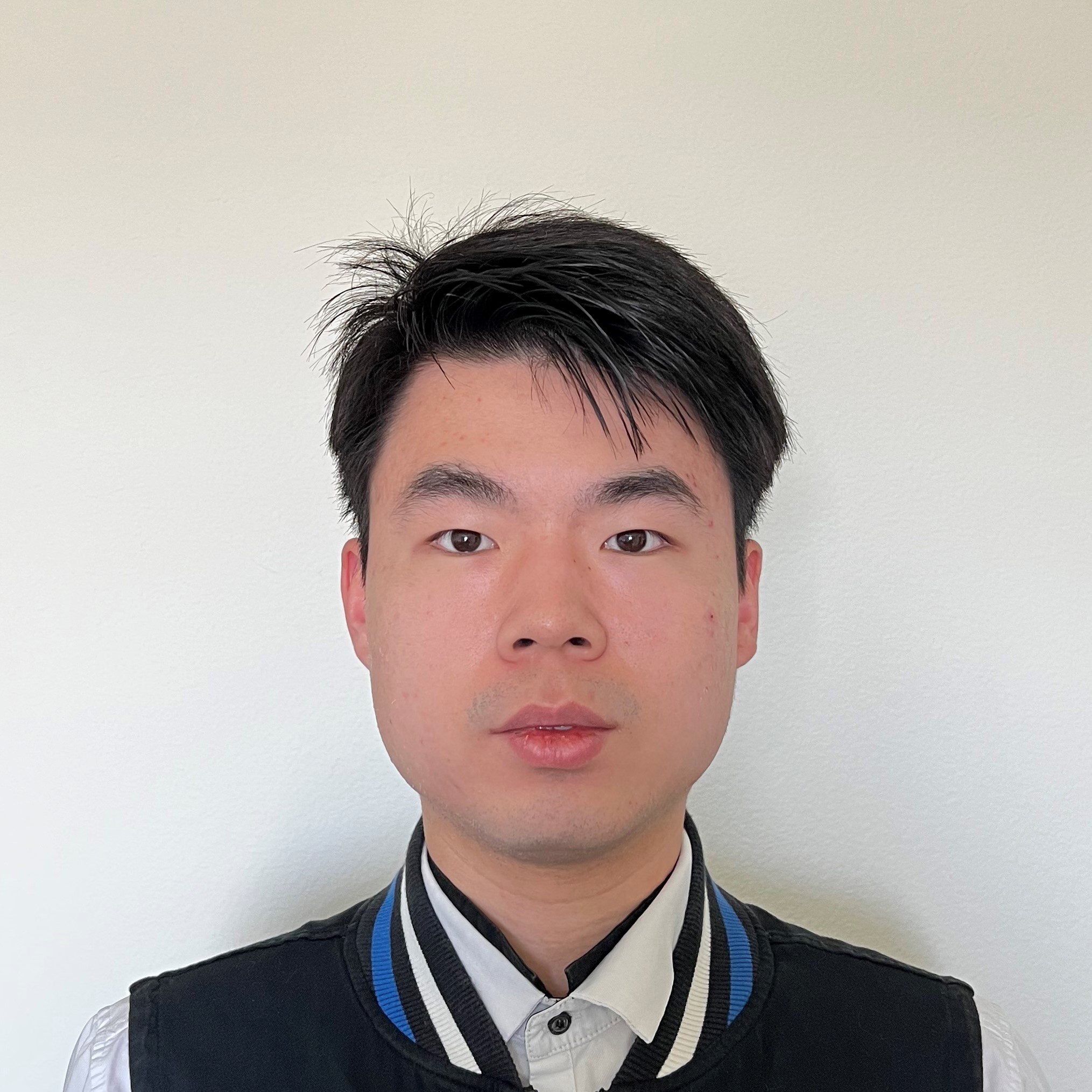}}]{Wang Hu}
	received the B.S. degree in electronics and information engineering from North China Electric Power University, Beijing, China. He is pursuing his Ph.D. degree in Electrical and Computer Engineering in University of California, Riverside, CA, USA. His research interests include automated driving, GNSS, aided inertial navigation, state estimation.
\end{IEEEbiography}

\begin{IEEEbiography}[{\includegraphics[width=1in,height=1.25in,clip,keepaspectratio]{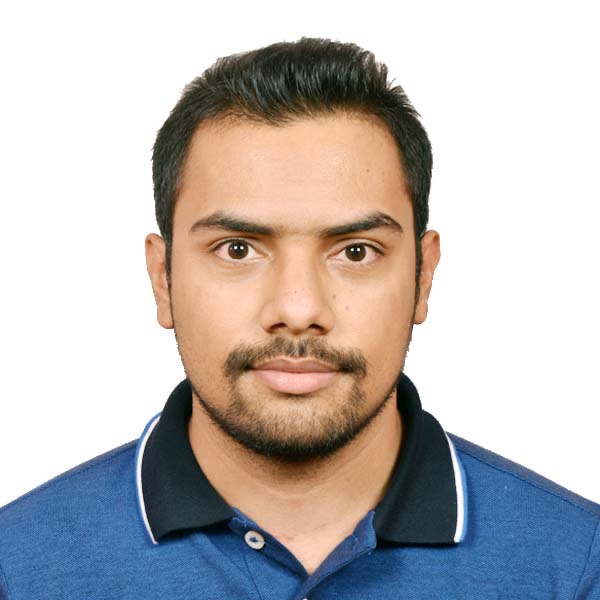}}]{Ashim Neupane}
	received the B.S. degree in mechanical engineering from Howard University, Washington, D.C., USA. He is pursuing his Ph.D. degree in Electrical and Computer Engineering in University of California, Riverside, CA, USA. His research interests include navigation, state estimation, robotics and controls.
\end{IEEEbiography}

\begin{IEEEbiography}[{\includegraphics[width=1in,height=1.25in,clip,keepaspectratio]{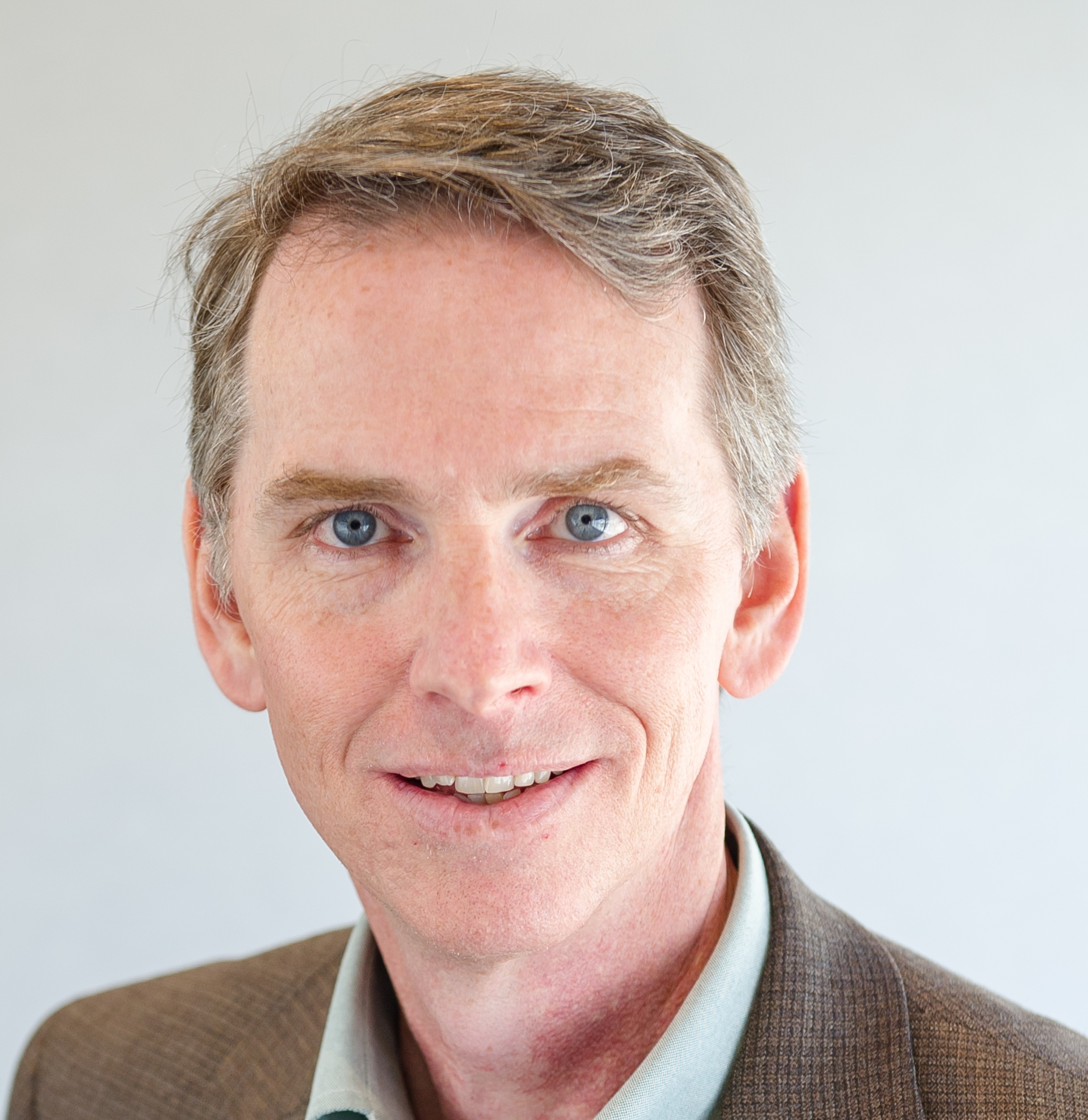}}]{Jay A. Farrell}
	received the B.S. degrees in physics and electrical engineering from Iowa State University, Ames, IA, USA, and the M.S. and Ph.D. degrees in electrical engineering from the University of Notre Dame, Notre Dame, IN, USA. After working at Charles Stark Draper Lab (1989–1994), he joined the University of California, Riverside, CA, USA, where he is the KA Endowed Chair in the Department of Electrical and Computer Engineering. He has authored over 250 technical articles and three books. 
	He was named a GNSS Leader to Watch for 2009–2010 by GPS World Magazine in May 2009.
	He was General Chair for IEEE CDC 2012 and has served the IEEE Control Systems Society and the American Association of Automatic Control as President. 
	Prof. Farrell is  a Distinguished Member of IEEE CSS and a Fellow of the IEEE, IFAC, and AAAS. 
\end{IEEEbiography}




%% file: main.bbl
\begin{thebibliography}{10}
\providecommand{\url}[1]{#1}
\csname url@samestyle\endcsname
\providecommand{\newblock}{\relax}
\providecommand{\bibinfo}[2]{#2}
\providecommand{\BIBentrySTDinterwordspacing}{\spaceskip=0pt\relax}
\providecommand{\BIBentryALTinterwordstretchfactor}{4}
\providecommand{\BIBentryALTinterwordspacing}{\spaceskip=\fontdimen2\font plus
\BIBentryALTinterwordstretchfactor\fontdimen3\font minus
  \fontdimen4\font\relax}
\providecommand{\BIBforeignlanguage}[2]{{%
\expandafter\ifx\csname l@#1\endcsname\relax
\typeout{** WARNING: IEEEtran.bst: No hyphenation pattern has been}%
\typeout{** loaded for the language `#1'. Using the pattern for}%
\typeout{** the default language instead.}%
\else
\language=\csname l@#1\endcsname
\fi
#2}}
\providecommand{\BIBdecl}{\relax}
\BIBdecl

\bibitem{sae2016}
``{On-Board System Requirements for {V2V} Safety Communications
  {J2945/1\_201603}},'' \emph{Society of Automotive Engineers}, 2016.

\bibitem{farrell2012its}
A.~Vu, A.~Ramanandan, A.~Chen, J.~A. Farrell, and M.~Barth, ``{Real-Time
  Computer Vision/DGPS-Aided Inertial Navigation System for Lane-Level Vehicle
  Navigation},'' \emph{{IEEE} Trans. Intell. Transp. Syst.}, vol.~13, no.~2,
  pp. 899--913, 2012.

\bibitem{NigelBarth2021}
N.~Williams and M.~Barth, ``{A qualitative analysis of vehicle positioning
  requirements for connected vehicle applications},'' \emph{{IEEE} Intell.
  Transp. Syst. Mag.}, vol.~13, no.~1, pp. 225--242, 2021.

\bibitem{barth2012eco}
K.~Boriboonsomsin, M.~J. Barth, W.~Zhu, and A.~Vu, ``Eco-routing navigation
  system based on multisource historical and real-time traffic information,''
  \emph{{IEEE} Trans. Intell. Transp. Syst.}, vol.~13, no.~4, pp. 1694--1704,
  2012.

\bibitem{team2014global}
G.~P. Team, ``{Global positioning system standard positioning service
  performance analysis report},'' \emph{GPS Product Team: Washington, DC, USA},
  2020.

\bibitem{beidou2018os}
``{BeiDou navigation satellite system open service performance standard
  (Version 2.0)},'' \emph{China Satellite Navigation Office}, 2018.

\bibitem{galileo2019os}
``{European GNSS (Galileo) open service service definition document},''
  \emph{European GNSS Service Centre}, 2019.

\bibitem{teunissen2017springer}
P.~Teunissen and O.~Montenbruck, \emph{{Springer handbook of global navigation
  satellite systems}}.\hskip 1em plus 0.5em minus 0.4em\relax Springer, 2017.

\bibitem{rahman2019ecef}
F.~Rahman, F.~Silva, Z.~Jiang, and J.~A. Farrell, ``{ECEF position accuracy and
  reliability: Continent scale differential GNSS approaches ({Phase C
  report})},'' \emph{https://escholarship.org/uc/item/05p9p3c9}, 2019.

\bibitem{chen2011trimble}
X.~Chen, T.~Allison, W.~Cao, K.~Ferguson, S.~Grunig, V.~Gomez, A.~Kipka,
  J.~Kohler, H.~Landau, R.~Leandro \emph{et~al.}, ``{Trimble RTX, an innovative
  new approach for network RTK},'' in \emph{Proc. of the 24th international
  technical meeting of the satellite division of the institute of navigation},
  2011, pp. 2214--2219.

\bibitem{vollath2000multi}
U.~Vollath, A.~Buecherl, H.~Landau, C.~Pagels, and B.~Wagner, ``{Multi-base RTK
  positioning using virtual reference stations},'' in \emph{Proc. of the 13th
  International Technical Meeting of the Satellite Division of The Institute of
  Navigation}, 2000, pp. 123--131.

\bibitem{wanninger1998real}
L.~Wanninger, ``{Real-time differential GPS error modelling in regional
  reference station networks},'' in \emph{Advances in positioning and reference
  frames}.\hskip 1em plus 0.5em minus 0.4em\relax Springer, 1998, pp. 86--92.

\bibitem{trimble_vrs}
{Trimble VRS Coverage},
  \url{https://positioningservices.trimble.com/resources/coverage-maps-3/}.

\bibitem{omni_vbs}
{OmniSTAR VBS},
  \url{https://www.omnistar.com/subscription-services/omnistar-vbs/}.

\bibitem{qianxun_findm}
{Qianxun SI FindM}, \url{https://www.qxwz.com/en/products}.

\bibitem{icd2013global}
``{Global positioning systems directorate system engineering \& integration
  interface specification IS-GPS-200H},'' \emph{Navstar GPS Space
  Segment/Navigation User Interfaces}, 2013.

\bibitem{galileo2008galileo}
``{Galileo signal in space interface control document ({OS SIS ICD})},''
  \emph{Galileo Open Service}, 2008.

\bibitem{beidou2012beidou}
``{BeiDou navigation satellite system signal in space interface control
  document open service signal {B1I} (Version 3.0)},'' \emph{China Satellite
  Navigation Office}, 2019.

\bibitem{glonass2008glonass}
``{Glonass interface control document},'' \emph{Russian Institute of Space
  Device Engineering: Moscow, Russia}, 2008.

\bibitem{song2014impact}
W.~Song, W.~Yi, Y.~Lou, C.~Shi, Y.~Yao, Y.~Liu, Y.~Mao, and Y.~Xiang, ``{Impact
  of GLONASS pseudorange inter-channel biases on satellite clock
  corrections},'' \emph{GPS Solutions}, vol.~18, no.~3, pp. 323--333, 2014.

\bibitem{wan2012carrier}
L.~Wanninger, ``{Carrier-phase inter-frequency biases of GLONASS receivers},''
  \emph{J. of Geodesy}, vol.~86, no.~2, pp. 139--148, 2012.

\bibitem{hu2019derivation}
W.~Hu and J.~A. Farrell, ``{Derivation of earth-rotation correction (Sagnac)
  and analysis of the effect of receiver clock bias},''
  \emph{https://escholarship.org/uc/item/1bf6w7j5}, 2019.

\bibitem{bryan2018eph}
B.~Stressler, J.~Heck, and H.~Steve, ``{An assessment of the accuracy of
  broadcast ephemerides for {multi-GNSS} positioning},'' in \emph{IGS
  workshop}, 2018.

\bibitem{geng2019modified}
J.~Geng, X.~Chen, Y.~Pan, and Q.~Zhao, ``{A modified lock/bias model to improve
  PPP ambiguity resolution at Wuhan University},'' \emph{J. of Geodesy},
  vol.~93, no.~10, pp. 2053--2067, 2019.

\bibitem{wang2016deter}
N.~Wang, Y.~Yuan, Z.~Li, O.~Montenbruck, and B.~Tan, ``{Determination of
  differential code biases with {multi-GNSS} observations},'' \emph{J. of
  Geodesy}, vol.~90, no.~3, pp. 209--228, 2016.

\bibitem{mont2014diff}
O.~Montenbruck, A.~Hauschild, and P.~Steigenberger, ``{Differential code bias
  estimation using {multi-GNSS} observations and global ionosphere maps},''
  \emph{J. of the Inst. of Navigation}, vol.~61, no.~3, pp. 191--201, 2014.

\bibitem{bock2001atm}
O.~Bock and E.~Doerflinger, ``{Atmospheric modeling in GPS data analysis for
  high accuracy positioning},'' \emph{Physics and Chemistry of the Earth, Part
  A}, vol.~26, no. 6-8, pp. 373--383, 2001.

\bibitem{farrell2008aided}
J.~A. Farrell, \emph{{Aided navigation: GPS with high rate sensors}}.\hskip 1em
  plus 0.5em minus 0.4em\relax McGraw-Hill, Inc., 2008.

\bibitem{laurichesse2010real}
D.~Laurichesse, F.~Mercier, and J.-P. Berthias, ``{Real-time PPP with
  undifferenced integer ambiguity resolution, experimental results},'' in
  \emph{Proc. of the 23rd International Technical Meeting of The Satellite
  Division of the Institute of Navigation}, 2010, pp. 2534--2544.

\bibitem{igs2020ssr}
``{IGS state space representation (SSR) format version 1.00},'' \emph{IGS
  Real-Time Working Group}, 2020.

\bibitem{hadas2015igs}
T.~Hadas and J.~Bosy, ``{IGS RTS precise orbits and clocks verification and
  quality degradation over time},'' \emph{GPS Solutions}, vol.~19, no.~1, pp.
  93--105, 2015.

\bibitem{sardon1994est}
E.~Sardon, A.~Rius, and N.~Zarraoa, ``{Estimation of the transmitter and
  receiver differential biases and the ionospheric total electron content from
  Global Positioning System observations},'' \emph{Radio Science}, vol.~29,
  no.~03, pp. 577--586, 1994.

\bibitem{ray2005geodetic}
J.~Ray and K.~Senior, ``{Geodetic techniques for time and frequency comparisons
  using GPS phase and code measurements},'' \emph{Metrologia}, vol.~42, no.~4,
  p. 215, 2005.

\bibitem{sch2016sinex}
S.~Schaer, ``{SINEX BIAS—Solution (Software/technique) independent exchange
  format for GNSS biases version 1.00},'' in \emph{IGS Workshop on GNSS Biases,
  Bern, Switzerland}, 2016.

\bibitem{wang2020gps}
N.~Wang, Z.~Li, B.~Duan, U.~Hugentobler, and L.~Wang, ``{GPS and GLONASS
  observable-specific code bias estimation: comparison of solutions from the
  IGS and MGEX networks},'' \emph{J. of Geodesy}, vol.~94, no.~8, pp. 1--15,
  2020.

\bibitem{leandro2006unb}
R.~Leandro, M.~Santos, and R.~Langley, ``Unb neutral atmosphere models:
  development and performance,'' in \emph{Proc. of the National Technical
  Meeting of The Institute of Navigation}, 2006, pp. 564--573.

\bibitem{li2012new}
W.~Li, Y.~Yuan, J.~Ou, H.~Li, and Z.~Li, ``{A new global zenith tropospheric
  delay model IGGtrop for GNSS applications},'' \emph{Chinese Science
  Bulletin}, vol.~57, no.~17, pp. 2132--2139, 2012.

\bibitem{kazm2017trop}
K.~Kazmierski, M.~Santos, and J.~Bosy, ``{Tropospheric delay modelling for the
  EGNOS augmentation system},'' \emph{Survey Review}, vol.~49, no. 357, pp.
  399--407, 2017.

\bibitem{Penna2001}
N.~Penna, A.~Dodson, and W.~Chen, ``{Assessment of EGNOS tropospheric
  correction model},'' \emph{J. of Navigation}, vol.~54, pp. 37 -- 55, 01 2001.

\bibitem{li2015new}
W.~Li, Y.~Yuan, J.~Ou, Y.~Chai, Z.~Li, Y.-A. Liou, and N.~Wang, ``{New versions
  of the BDS/GNSS zenith tropospheric delay model IGGtrop},'' \emph{J. of
  Geodesy}, vol.~89, no.~1, pp. 73--80, 2015.

\bibitem{li2018}
W.~Li, Y.~Yuan, J.~Ou, and Y.~He, ``{IGGtrop\_SH and IGGtrop\_rH: Two improved
  empirical tropospheric delay models based on vertical reduction functions},''
  \emph{{IEEE} Trans. Geosci. Remote Sens.}, vol.~56, no.~9, pp. 5276--5288,
  2018.

\bibitem{klobuchar1987}
J.~A. Klobuchar, ``{Ionospheric time-delay algorithm for single-frequency GPS
  users},'' \emph{{IEEE} Trans. Aerosp. Electron. Syst.}, no.~3, pp. 325--331,
  1987.

\bibitem{gal2020iono}
``{Ionospheric correction algorithm for Galileo single frequency users
  (V1.2)},'' \emph{European GNSS Service Centre}, 2016.

\bibitem{yuan2008refi}
Y.~Yuan, X.~Huo, J.~Ou, K.~Zhang, Y.~Chai, D.~Wen, and R.~Grenfell, ``{Refining
  the Klobuchar ionospheric coefficients based on GPS observations},''
  \emph{{IEEE} Trans. Aerosp. Electron. Syst.}, vol.~44, no.~4, pp. 1498--1510,
  2008.

\bibitem{angrisano2013ass}
A.~Angrisano, S.~Gaglione, C.~Gioia, M.~Massaro, and U.~Robustelli,
  ``{Assessment of NeQuick ionospheric model for Galileo single-frequency
  users},'' \emph{Acta Geophysica}, vol.~61, no.~6, pp. 1457--1476, 2013.

\bibitem{rtcm2014proposal}
R.~S. Committee \emph{et~al.}, ``{Proposal of new RTCM SSR messages, SSR stage
  2: Vertical TEC (VTEC) for RTCM standard 10403.2 differential GNSS (Global
  Navigation Satellite Systems) services-version 3},'' \emph{RTCM Special
  Committee}, no. 104, 2014.

\bibitem{li2020igs}
Z.~Li, N.~Wang, M.~Hernandez-Pajares, Y.~Yuan, A.~Krankowski, A.~Liu, J.~Zha,
  A.~Garcia-Rigo, D.~Roma-Dollase, H.~Yang \emph{et~al.}, ``{IGS real-time
  service for global ionospheric total electron content modeling},'' \emph{J.
  of Geodesy}, vol.~94, no.~3, pp. 1--16, 2020.

\bibitem{li2015shpts}
Z.~Li, Y.~Yuan, N.~Wang, M.~Hernandez-Pajares, and X.~Huo, ``{SHPTS: Towards a
  new method for generating precise global ionospheric TEC map based on
  spherical harmonic and generalized trigonometric series functions},''
  \emph{J. of Geodesy}, vol.~89, no.~4, pp. 331--345, 2015.

\bibitem{cne2015vtec}
D.~Laurichesse and A.~Blot, \emph{{New CNES real time products including
  BeiDou}}.\hskip 1em plus 0.5em minus 0.4em\relax IGS Mail No. 7183, 10 Nov
  2015.

\bibitem{fuller2005ustec}
T.~Fuller-Rowell, ``{USTEC: a new product from the space environment center
  characterizing the ionospheric total electron content},'' \emph{GPS
  Solutions}, vol.~9, no.~3, pp. 236--239, 2005.

\bibitem{ustecweb}
{USTEC Real-time TEC},
  \url{https://services.swpc.noaa.gov/text/us-tec-total-electron-content.txt}.

\bibitem{el1993faa}
M.~B. El-Arini, P.~A. O'Donnell, P.~M. Kellam, J.~A. Klobachar, T.~C. Wisser,
  and P.~H. Doherty, ``{The FAA wide area differential GPS (WADGPS) static
  ionospheric experiment},'' in \emph{Proc. of the National Technical Meeting
  of the Institute of Navigation}, 1993, pp. 485--496.

\bibitem{prol2017comparative}
F.~D.~S. Prol, P.~D.~O. Camargo, and M.~T. D. A.~H. Muella, ``{Comparative
  study of methods for calculating ionospheric points and describing the GNSS
  signal path},'' \emph{Boletim de Ci{\^e}ncias Geod{\'e}sicas}, vol.~23,
  no.~4, pp. 669--683, 2017.

\bibitem{PsiakiMohiuddin2012}
M.~L. Psiaki and S.~Mohiuddin, ``{Modeling, analysis, and simulaiton of GPS
  carrier phase for spacecraft relative navigation},'' \emph{J. of Guidance,
  Control, and Dynamics}, vol.~30, no.~6, pp. 1682--1639, 2007.

\bibitem{weber2016bkg}
G.~Weber, L.~Mervart, and A.~St{\"u}rze, \emph{{BKG Ntrip Client (BNC): Version
  2.12}}.\hskip 1em plus 0.5em minus 0.4em\relax Verlag des Bundesamtes f{\"u}r
  Kartographie und Geod{\"a}sie, 2016.

\bibitem{rtklibgit}
T.~Takasu and A.~Yasuda, {RTKLIB Repository:}
  \url{https://github.com/tomojitakasu/RTKLIB}.

\bibitem{takasu2009development}
{T. Takasu and A. Yasuda}, ``{Development of the low-cost RTK-GPS receiver with
  an open source program package RTKLIB},'' \emph{Int. Symposium on GPS/GNSS},
  vol.~1, 2009.

\bibitem{pppwizard}
{The PPP-Wizard project of CNES}, \url{http://www.ppp-wizard.net/news.html}.

\bibitem{wang2018assessment}
Z.~Wang, Z.~Li, L.~Wang, X.~Wang, and H.~Yuan, ``{Assessment of multiple GNSS
  real-time SSR products from different analysis centers},'' \emph{ISPRS Int.
  Journal of Geo-information}, vol.~7, no.~3, p.~85, 2018.

\bibitem{wang2018code}
N.~Wang, Z.~Li, and Y.~Yuan, ``{Multi-GNSS code bias handling: an observation
  specific perspective},'' in \emph{IGS workshop}, 2018.

\bibitem{cai2011unmanned}
G.~Cai, B.~M. Chen, and T.~H. Lee, \emph{{Unmanned rotorcraft systems}}.\hskip
  1em plus 0.5em minus 0.4em\relax Springer Science \& Business Media, 2011.

\bibitem{waas2020report}
``{Wide area augmentation system performance analysis report},''
  \emph{nstb.tc.faa.gov}, 2020.

\bibitem{geng2019pride}
J.~Geng, X.~Chen, Y.~Pan, S.~Mao, C.~Li, J.~Zhou, and K.~Zhang, ``{PRIDE
  PPP-AR: An open-source software for GPS PPP ambiguity resolution},''
  \emph{GPS Solutions}, vol.~23, no.~4, p.~91, 2019.

\bibitem{li2018multi}
X.~Li, X.~Li, Y.~Yuan, K.~Zhang, X.~Zhang, and J.~Wickert, ``{Multi-GNSS phase
  delay estimation and PPP ambiguity resolution: GPS, BDS, GLONASS, Galileo},''
  \emph{J. of Geodesy}, vol.~92, no.~6, pp. 579--608, 2018.

\end{thebibliography}
